\documentclass[%
 amsmath,amssymb,
 aps,
]{revtex4-2}

\usepackage{graphicx}
\usepackage{dcolumn}
\usepackage{bm}

\usepackage{tikz}
\usepackage[dvipsnames]{xcolor}
\usetikzlibrary{positioning}
\usepackage{mwe}
\usepackage{natbib}
\usepackage{hyperref}
\usepackage{physics}
\usepackage{comment}
\usepackage{multirow}
\usepackage[utf8]{inputenc}
\usepackage[T1]{fontenc}
\usepackage{mathptmx}
\usepackage{etoolbox}
\usepackage{booktabs}
\usepackage{subfigure}

\makeatletter
\def\@email#1#2{%
 \endgroup
 \patchcmd{\titleblock@produce}
  {\frontmatter@RRAPformat}
  {\frontmatter@RRAPformat{\produce@RRAP{*#1\href{mailto:#2}{#2}}}\frontmatter@RRAPformat}
  {}{}
}%
\makeatother
\begin{document}

\preprint{APS/123-QED}

\title[Dipolar solvent contributions for transient nanoscale EOF]{Dipolar solvent contributions for transient nanoscale electroosmotic flow}
\author{Pramodt Srinivasula}
 \altaffiliation[Previously at ]{the Department of Mechanical Engineering, TU Darmstadt, Germany, where part of the research was conducted.}
 \email{pramodt.research@gmail.com}
\affiliation{ 
ElectroSoft Labs LLP, Mumbai, India
}%


\date{\today}

\begin{abstract}
Electrohydrodynamic flows of electrolytes with low to moderate ion concentration at the nanoscale are significantly influenced by the molecular structure of water-like polar solvents within the electric double layer (EDL). Moreover, unlike in microfluidics, at these length scales the time scale of evolution of EDL often becomes comparable to the consequent fluidic phenomena of interest. 
While continuum descriptions to model such phenomena typically assume a constant dielectric and viscous solvent background, this study incorporates dipolar solvent physics. Specifically, both dielectric saturation and the viscoelectric effect are implemented together into a Poisson-Nernst-Planck-Stokes framework, using the Langevin-Bikerman solvent permittivity distribution and empirical viscoelectric coefficients, respectively.
Numerical simulations in a one-dimensional geometry reveal substantial modifications to the electrohydrodynamic body force density and transient electroosmotic mobility during EDL evolution. The magnitude and temporal evolution of these corrections are characterized across parametric regimes, revealing systematic departures from standard constant-permittivity and constant-viscosity models, with electroosmotic mobility reductions of up to $65\%$ governed by a characteristic dimensionless parameter. The results predict a characteristic frequency-dependent transition in electroosmotic mobility and dipolar solvent corrections in emerging MHz AC electrokinetic flows.
The results provide a solvent-consistent continuum framework for transient nanoscale electroosmotic flows and quantify the impact of molecular solvent structure on electrohydrodynamic transport relevant to modern nanofluidic applications.
\end{abstract}

\maketitle
\section{Introduction}
Recent advances in nanofluidic and nanopore-based technologies, most prominently in ionic logic devices \citep{wang2025field}, DNA sequencing \citep{melnikov2024ionic}, energy harvesting \citep{dolatshahi2025salinity} and ultra-high frequency nanoscale electroosmotic flow \citep{blankenburg2025nanoscale} have brought renewed attention to electrohydrodynamic (EHD) transport under nanoconfinement. At these length scales, transport is governed by the formation and evolution of electric double layers (EDLs), whose characteristic spatial and temporal scales become comparable to those accessed during device operation.
Consequently, transient electrokinetic phenomena play an increasingly important role in determining transport behavior. Moreover, unlike conventional microfluidic systems, the strong electric fields and molecular confinement within nanoscale EDLs render solvent-dependent effects, particularly dielectric saturation and viscoelectricity in low-to-moderate concentration aqueous electrolytes, increasingly significant. These developments motivate the need for continuum formulations that consistently incorporate molecular solvent physics into transient electrokinetic analyses.

Time-dependent EDL dynamics play a significant role in modern nanofluidic applications, particularly nanopore-based biosensing and sequencing, where macromolecules and nanoparticles are identified from transient ionic current signatures generated during their pore translocation \citep{he2011gate,he2011controlling,wei2024nanopore}. Recent experimental advances have achieved sub-microsecond temporal resolution to capture these rapidly evolving current signals \citep{shekar2016measurement,rosenstein2012integrated}. The measured ionic current reflects the coupled evolution of ionic charge redistribution in the electric double layer (EDL), and electroosmotic flow as the particle traverses the nanopore. Consequently, physically consistent models of transient EDL dynamics are important for interpreting high-bandwidth nanopore measurements. Similar transient electrokinetic phenomena also arise in shock electrodialysis \citep{tian2021theory}, electroconvective transport in nanoporous membranes \citep{de2018confined,choi2024electroconvective}, and nanofluidic energy-conversion \citep{chapuis2025boosting} and separation devices \citep{abdelghani2025resonant}, motivating physically consistent models of transient ion transport and electrohydrodynamic flow.

Recent advances in high-frequency AC measurements of electric double layer (EDL) and electroosmotic flow (EOF) dynamics are enabling new investigations in AC electrokinetics \citep{shiddiky2014molecular} and electrochemical energy-storage systems such as supercapacitors \citep{li2025high,maurel2022operando}.  Recently, \citet{blankenburg2025nanoscale} experimentally demonstrated AC-driven electroosmotic flow at frequencies up to 100 MHz, including the observation of flow reversal under high-frequency forcing. This emerging direction is further supported by recent theoretical \citep{coquinot2026electron} and molecular dynamics studies \citep{alosious2026study} of high-frequency AC electrokinetic flows, together indicating significant transient electrohydrodynamic coupling beyond classical charge-relaxation limits. This provokes the interest in more careful estimation of nanoscale transient EOF, which could differ significantly from that at microscale.

Beyond the limited experimental studies on AC electroosmotic flow modulation in asymmetric nanopores \citep{blankenburg2025nanoscale, lan2016voltage}, direct measurements of transient nanoscale EOF remain scarce. In contrast, transient electric double layer (EDL) charging has been investigated using electrolytic cell configurations. The seminal work of Bazant \textit{et al.} \citep{bazant2004diffuse} analyzed the diffuse-charge dynamics between two planar electrodes subjected to a step potential difference using a one-dimensional continuum model, establishing a cornerstone framework for transient electrochemical transport. Extending this paradigm to nanoconfinement, Tivony \textit{et al.} \citep{tivony2018charging} experimentally investigated transient EDL charging in a tunable nanoslit formed between a gold surface and a mica surface using a surface force balance (SFB), providing direct insight into charging dynamics relevant to nanopores. Although the experiments measured transient EDL relaxation within a two-wall nanoslit following a potential step, the authors interpreted the observed charging times as nanopore charging by invoking the transmission-line model, thereby attributing the dynamics to reservoir-to-pore ion transport without explicitly modeling that transport. This demonstrates the usefulness of fundamental learning of the transient charging phenomena from such an electrolytic cell setup to nanoscale domains beyond. Accordingly, this canonical electrolytic cell configuration has become a well-established benchmark for comparing transient response predictions from the classical PNP model \citep{bazant2004diffuse}, modified PNP formulations \citep{kilic2007bsteric}, and other possible enriched continuum frameworks.

Apart from a few molecular dynamics studies \citep{cui2004electroosmotic}, however, continuum analyses of transient electroosmotic flow in this canonical configuration remain largely unexplored. The multiscale nature of these systems, characterized by nanometer confinement, submicron device dimensions, and above nanosecond transient dynamics, renders direct molecular simulations computationally prohibitive. The underlying electrokinetic phenomena in such studies are commonly modeled using a one-dimensional continuum formulation, which adequately captures the short-time charging dynamics at nanoscales and provides an accurate approximation for higher-dimensional systems during the early transient regime \citep{palaia2025charging}.  This highlights the need for predictive continuum analysis capturing the coupled transient evolution of ion transport and electrohydrodynamic flow under high-frequency forcing, extending beyond the well-established steady-state electrokinetic formulations \citep{peng2025ionic}.

At nanometer length scales, electrokinetic transport increasingly departs from the assumptions underlying classical continuum electrokinetic transport \citep{pennathur2005electrokinetic}. Consequently, predictive continuum models have progressively incorporated molecular-scale corrections to extend their range of validity \citep{kavokine2021fluids,di2025validity}. Steric effects are commonly represented using lattice-gas formulations, such as the Bikerman model \citep{borukhov1997steric,kilic2007bsteric}, or interacting hard-sphere theories such as BMCSL \citep{pandey2021impact}. Ion--solvent interactions have likewise been incorporated through dielectrophoretic forces on solvated ions and Born energy corrections \citep{pandey2021impact}. These effects generally enhance electroosmotic flow (EOF), whereas electroviscous and viscoelectric effects suppress it \citep{pandey2021impact,pandey2022effects,sahu2024ion}. More computationally intensive approaches, including immersed-particle methods, have also been developed to explicitly capture discrete ion--solvent interactions \citep{ladiges2021discrete}. Despite these advances, most continuum formulations assume a spatially uniform solvent permittivity, thereby neglecting the dipolar polarization of the solvent itself. This approximation becomes increasingly restrictive in low- and moderately concentrated aqueous electrolytes, where solvent polarization significantly influences nanoscale electrokinetic transport.

Among the molecular-scale effects, the dipolar polarization of the solvent plays a fundamental role in determining both the dielectric and hydrodynamic response of aqueous electrolytes at the nanoscale. Direct measurements have revealed a pronounced reduction in the effective dielectric permittivity of nanoconfined water \citep{fumagalli2018anomalously}, while molecular dynamics simulations have demonstrated significant modifications of solvent polarization near charged interfaces \citep{zheng2006surfaces}. Both nanoconfinement and the strong electric fields within the electric double layer restrict the orientational response of water molecular dipoles, thereby reducing the effective dielectric permittivity. Continuum formulations incorporating dipolar solvent physics, most notably the Langevin--Bikerman (LB) and related Landau-type models \citep{iglivc2010excluded,monet2021nonlocal}, successfully capture the dielectric saturation and excluded-volume effects in the electric double layer (EDL). These models predict significant modifications to the electrostatic potential, ion distributions, and electrostatic pressure at steady-state \citep{abrashkin2007dipolar,iglivc2010excluded}. 

Transient electroosmotic flow at the EDL charging timescale has been investigated in the context of emerging applications such as resonant nanofluidic pumps \citep{ratschow2022resonant} and convection-enhanced nanofluidic capacitors \citep{ratschow2025convection}. These studies, however, rely on simplified continuum formulations without accounting for dipolar solvent effects. On the other hand, several recent studies have incorporated dipolar solvent polarization into generalized Poisson--Nernst--Planck (PNP) formulations for transient modeling through field-dependent polarization densities and dielectric permittivities \citep{liu2018modified,liu2020molecular,gui2024ion}. However, while these formulations provide a rigorous description of solvent polarization, they typically introduce higher-order, strongly nonlinear governing equations that complicate numerical implementation and coupling with hydrodynamic solvers. Consequently, their applications have largely been confined to steady-state electrostatic and ion-transport problems, with only limited extensions to steady-state electroosmotic flow \citep{saurabh2023mathematical}. The development of robust continuum formulations capable of efficiently incorporating dipolar solvent physics into transient electroosmotic flow and electrohydrodynamic stress analyses remains largely unexplored.

A parallel but largely independent body of works in the literature concerns the viscoelectric effect, the increase of liquid viscosity under strong electric fields due to dipolar reorientation. Originally identified in polar organic liquids \cite{andrade1946effect,andrade1951effect} and later extrapolated to water \cite{lyklema1961interpretation, hunter1978viscoelectric}, the viscoelectric effect is commonly modeled through an empirical quadratic dependence of viscosity on electric field strength. Recent surface-force-balance experiments have directly confirmed the magnitude of this effect in water \citep{jin2022direct}, lending experimental support to decades-old theoretical estimates. Nevertheless, most electrokinetic models treat viscoelectricity as a secondary near-wall correction and rarely consider it together with dielectric saturation or transient EDL dynamics. At nanometer confinement, where electric fields are strong and electroosmotic flow is closely coupled to EDL dynamics, a unified continuum description incorporating these effects becomes increasingly desirable.

Electroosmotic mobility in nanochannels is governed by the coupled evolution of ionic charge, solvent polarization, and viscous stresses within the electric double layer (EDL). Steady-state studies have consistently shown that dielectric saturation modifies electroosmotic mobility, including continuum formulations based on the Langevin--Bikerman (LB) framework \citep{sinha2016effect,sin2018influence,srinivasula2025dipolar}; whereas viscoelectric effects generally suppress the flow, with important implications for nanoscale electrokinetic transport \citep{hsu2016electrokinetics,saurabh2023mathematical,mehta2023viscoelectric,mehta2025arresting}. Collectively, these studies demonstrate that continuum electrohydrodynamics remains predictive at the nanoscale when appropriate molecular-scale solvent physics is incorporated. However, analogous continuum frameworks for transient electroosmotic transport that simultaneously account for dielectric saturation, viscoelectricity, and coupled electrohydrodynamics remain largely unexplored.

This work develops a transient Poisson--Nernst--Planck--Stokes framework that incorporates dipolar solvent physics through dielectric saturation and viscoelectric effects. The canonical electrolytic cell configuration is adopted as a fundamental benchmark, in which electroosmotic flow is generated by a weak lateral electric field following a step-voltage excitation. Within this framework, the influence of dipolar solvent physics on transient EDL charging and the resulting electroosmotic flow under nanoscale confinement is systematically investigated. The derived transient scaling laws are further interpreted in the frequency domain to establish their implications for AC electrokinetic flows, providing generic insight into emerging high-frequency electrokinetic systems and ultrafast nanofluidic transport.

\section{Methodology}

\begin{figure} 
\centering
\begin{tikzpicture}
\sbox0{\includegraphics[width=0.65\textwidth]{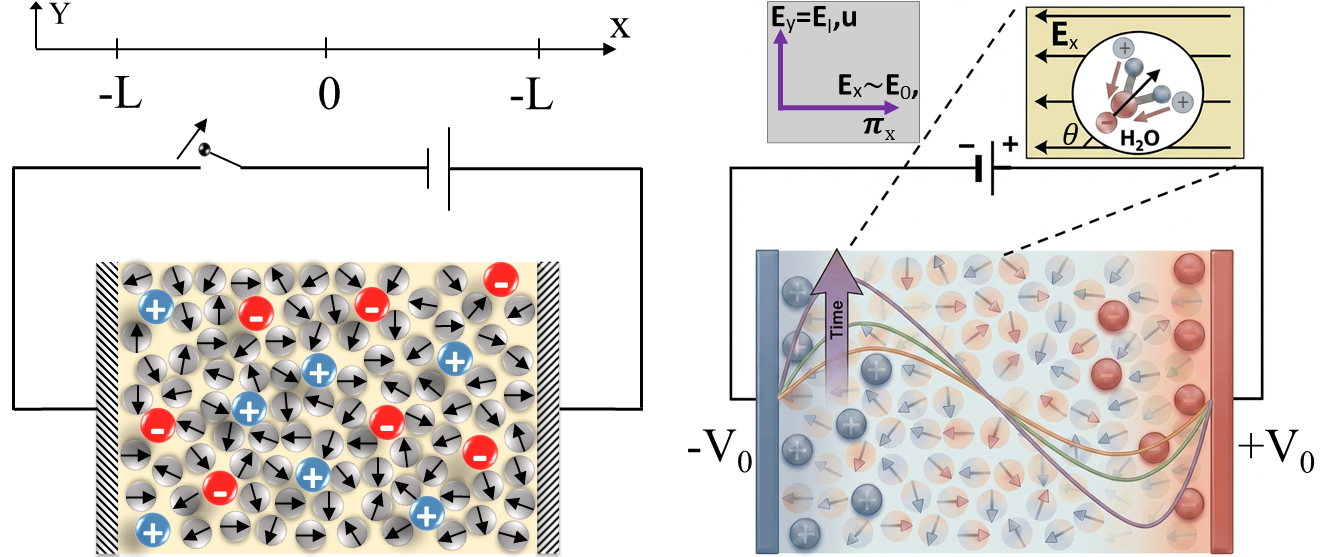} } 
  \node[above right,inner sep=0pt] at (0,0)  {\usebox{0}};
  \node[black] at (0.0\wd0,1.0\ht0) {(a)};
  \node[black] at (0.56\wd0,1.0\ht0) {(b)};
\end{tikzpicture}
\caption{\textit{Electrokinetic configuration and transient electroosmotic response.}
(a) A segment of the infinitely long electrolytic cell (along the y-direction) before the application of a longitudinal potential difference between the electrodes (along the x-direction). (b) Transient evolution of the electroosmotic velocity profile, $u(x,t)$ under a small lateral electric field $E_l$, and a longitudinal field of nominal magnitude $E_0\!=\!2V_0/L$) during electric double layer charging upon applying the potential difference between the electrodes at time $t\!=\!0$. Solvent polarization, characterized by the orientation angle $\theta(\mathbf{E})$ of the molecular dipoles, is incorporated in the asymptotic limit $E_l/E_0\ll1$.} \label{fig:1_illustration}
\end{figure}

A symmetric monovalent electrolyte with equal ionic diffusivities is confined between two large parallel planar electrodes separated by a distance $2L$, following the classical transient charging configuration of \citet{bazant2004diffuse,kilic2007bsteric}, as illustrated in Fig.~\ref{fig:1_illustration}(a,b). The electrolyte has bulk ionic number concentrations $n_\pm=n_0$ and is dissolved in water of molecular density $n_{0s}=55\,000$ mM within the incompressible limits. The solvent dipolar response is represented using an effective molecular dipole moment of $p_0\approx4.7$--$5$ D accounting for intermolecular interactions, corresponding to the zero-field relative permittivity $\varepsilon_{r0} =1 + n_{0s}p_0^2\beta/(3\varepsilon_0)$, with inverse temperature parameter $\beta=1/(k_B T)$ and permittivity of free space $\varepsilon_0$ \citep{abrashkin2007dipolar,iglivc2010excluded}.

The present formulation adopts a continuum mean-field description of a weakly to moderately concentrated aqueous electrolyte based on the Langevin--Bikerman (LB) free-energy framework \citep{kilic2007bsteric,iglivc2010excluded,gongadze2013spatial}. Besides the conventional electrostatic field energy, the LB model incorporates field--dipole interactions within a lattice-gas description of the solvent, giving rise to a field-dependent relative permittivity under strong electric fields. Additional energetic contributions, including spatial variations in ion Born energies and dipole self-energies, are neglected because they are expected to remain weak for the low to mild electrolyte concentrations considered here. Molecular-scale interfacial effects are represented through compact Stern layers adjacent to the electrode surfaces, providing an effective continuum description of dielectric screening and molecular adsorption near insulating or weakly reactive interfaces.

Building on these assumptions, a one-dimensional Poisson--Nernst--Planck (PNP)--Stokes formulation is employed to describe the coupled transient evolution of ion transport, dipolar solvent polarization, and electroosmotic flow.
At time $t=0$, a step potential difference of magnitude $2V_0$ is suddenly applied across the electrodes [Fig.~\ref{fig:1_illustration}(b)], initiating transient electric double layer (EDL) charging. The EDL dynamics are resolved along the direction normal to the electrodes ($x\in(-L,L)$), while electroosmotic flow is driven parallel to the electrode surfaces by a simultaneously imposed uniform lateral electric field $E_l$. The electric field normal to the electrodes is obtained from the local electric potential $\psi$ as, $E_x=-\partial\psi/\partial x$, is characterized by the nominal magnitude $E_0=V_0/L$, which serves as the reference field throughout the analysis. The imposed lateral field satisfies $E_l/E_0\ll1$, ensuring that the electrostatic charging process remains governed by the longitudinal field while the electroosmotic flow acts as a weak transverse response.

The electrode surfaces are assumed to be ideally insulating, with no Faradaic reactions or charge injection. Compact Stern layers of thickness $\lambda_s$ are introduced adjacent to the electrode surfaces ($x=\pm L$). Following \citet{bazant2004diffuse}, the transient calculations are initialized with a uniform electric field,
\begin{equation}
\psi(x,t\!=\!0)=E_0x,
\qquad
x\in[-L,L],
\label{eq:initial_psi}
\end{equation}
while the free ionic species are initially distributed uniformly throughout the electrolyte outside the Stern layers,
\begin{equation}
n_\pm(x,t\!=\!0)=n_0,
\qquad
x\in[-L+\lambda_s,L-\lambda_s].
\label{eq:initial_species}
\end{equation}

\subsection{Electrostatics}

Within the electrolyte, unlike the classical Poisson--Boltzmann framework, which assumes a spatially uniform solvent permittivity, the Langevin--Bikerman (LB) model explicitly accounts for the orientational polarization of solvent dipoles within a lattice-gas framework \citep{iglivc2010excluded}. Consequently, the local relative permittivity, $\varepsilon_r(x)$, becomes dependent on the electric-field strength, enabling the model to capture dielectric saturation under the strong electric fields characteristic of nanoconfined electric double layers. For computational efficiency, the spatial variation of $\varepsilon_r(x)$ is represented using an exponential approximation fitted to the steady-state LB solution of \citet{iglivc2010excluded}, with fitting parameters $A_1$--$A_3$ given in Appendix~A. The space charge density is $\rho_i(x,t)=e_0(n_+-n_-)$, with the elementary charge $e_0$. The governing Poisson equation for the electric potential $\psi(x,t)$ is therefore written as
\begin{align}
- \varepsilon_0 \pdv{x}\left(\varepsilon_r(x) \pdv{\psi}{x}\right)
&=
\rho_i,
\qquad
\varepsilon_r(x)
=
A_1 \varepsilon_{r0} 
\left(
1-
e^{-A_2\left(\frac{x-\lambda_s}{L}+1\right)} \right)
+
A_3\varepsilon_{r0} 
,
\qquad
x\in(-L+\lambda_s,L-\lambda_s),
\label{eq:dimen_Poisson}
\end{align}
where the decrease in $\varepsilon_r$ near the charged surface reflects dielectric saturation induced by the strong local electric field.

Within the Stern layer at the electrode surfaces, the relative permittivity $\varepsilon_{r,s}$ is prescribed as a uniform value lower than that of the bulk electrolyte to represent the reduced dielectric response arising from interfacial solvent ordering and specifically adsorbed ions. Assuming the absence of free charge within the Stern layer, the electric potential satisfies the Laplace equation,
\begin{align}
-\varepsilon_0\varepsilon_{r,s} \pdv[2]{\psi}{x}
&=
0,
\qquad
x\in\pm(L-\lambda_s,L),
\end{align}
subject to the applied electrode potential
\begin{align}
\psi=\pm V_0,
\qquad
x=\pm L.
\label{eq:SternBC}
\end{align}
The Stern-layer thickness is taken to be of the order of the solvent molecular size, $\lambda_s=a/2$, where $a$ is the effective diameter of the solvated ions (lattice spacing).

The relative permittivity (dielectric constant) of the Stern layer is influenced by several interrelated interfacial factors, including solvent molecular orientation, specifically adsorbed solvated ions, local electric-field strength, and surface charge density. Consequently, its value is generally system dependent and cannot be prescribed universally. In the present study, the applied surface potential (or surface charge) is fixed, and the Stern-layer relative permittivity is prescribed as $70\%$ of the bulk electrolyte value $\varepsilon_{r0}$, following the previous work \citep{srinivasula2025dipolar}. This assumption fixes the  electrode--electrolyte interfacial dielectric response, allowing the influence of dipolar solvent physics in the electrolyte, namely dielectric saturation and viscoelectricity, to be isolated and examined systematically.

Ionic transport is described by the modified Nernst--Planck equations, in which finite ion-size (steric) effects are incorporated through the lattice-gas entropy formulation of \citet{kilic2007bsteric}. For a monovalent electrolyte, the evolution of the positive and negative ion concentrations, $n_\pm(x,t)$, is governed by
\begin{align}
\pdv{n_\pm}{t}
=
-\pdv{J_\pm}{x}
=
D\left[
\pdv[2]{n_\pm}{x}
\pm
\beta e_0
\pdv{x}(n_\pm\nabla\psi)
+
a^3
\pdv{x}\left(
\frac{n_\pm \pdv{x}(n_++n_-)}
{1-a^3(n_++n_-)}
\right)
\right],
\label{eq:ModifNP}
\end{align}
where $J_\pm$ denote the ionic fluxes and $D$ is the characteristic diffusion coefficient of the ions. No-flux boundary conditions are imposed at the insulating surfaces,
\begin{align}
J_\pm=0,
\qquad
x=\pm L.
\label{eq:noFluxBC}
\end{align}

Throughout the transient evolution, solvent polarization is assumed to equilibrate instantaneously relative to the ionic timescale and therefore enters only through the field-dependent permittivity in Eq.~(\ref{eq:dimen_Poisson}). Together, the above Poisson and Nernst--Planck equations (Eqs. \ref{eq:dimen_Poisson}and \ref{eq:ModifNP}) constitute the Langevin--Bikerman-fitted transient (LBFT) electrostatic model employed throughout the present study. Besides that, two conventional continuum formulations can be identified as: the classical PNP--Stokes model ($\varepsilon_r=\varepsilon_{r0}$ in Eq.~\ref{eq:dimen_Poisson} and $a=0$ in Eq.~\ref{eq:ModifNP}) and the modified PNP (MPNP)--Stokes model ($\varepsilon_r=\varepsilon_{r0}$ in Eq.~\ref{eq:dimen_Poisson} and $a\neq0$ in Eq.~\ref{eq:ModifNP}).

The electrohydrodynamic response can be characterized by the surface force acting on the electrodes, evaluated from the normal component of the electrohydrodynamic (EHD) force density, $\Pi_x$ \citep{perkin2006long, abrashkin2007dipolar}. Here, $\Pi_x$ is associated with the longitudinal electric field $E_x\!=\!-\partial\psi/\partial x$, whereas the electroosmotic flow is driven by the imposed lateral electric field $E_l$.
For a dielectric liquid with spatially varying polarization, the EHD force density within the EDL is expressed as the sum of Coulombic and dielectrophoretic contributions \citep{saville1997electrohydrodynamics},
\begin{align}
\Pi_x
=
-\rho_i\pdv{\psi}{x}
-
P_x\pdv[2]{\psi}{x},
\end{align}
where $P_x$ is the polarization density normal to the electrode surfaces. In the present formulation, the polarization evolves self-consistently with the field-dependent permittivity through the Langevin--Bikerman model \citep{gongadze2013spatial}. The polarization density is therefore written as
\begin{align}
P_x = -p_0c_{0s}\, \mathcal{L}(\beta p_0E_x), \qquad
\mathcal{L}(\beta p_0E_x)=
\coth(\beta p_0E_x)
-
\frac{1}{\beta p_0E_x},
\end{align}
where $\mathcal{L}$ denotes the Langevin function.

\subsection{Hydrodynamics}

Application of a uniform lateral electric field $E_l$ parallel to the charged surfaces, as illustrated in Fig.~\ref{fig:1_illustration}(c), exerts an electrostatic body force $\rho_i(x,t)E_l$ on the net ionic charge residing within the transient electric double layer (EDL), thereby driving electroosmotic flow parallel to the walls. Since the flow is unidirectional and varies only across the channel thickness, momentum diffuses over the characteristic distance $L\sim10\,\mathrm{nm}$ between the charged surfaces on the viscous timescale, $\tau_h=L^2/\nu$, which is approximately $0.1\,\mathrm{ns}$ for aqueous electrolytes (for the kinematic viscosity $\nu\approx10^{-6}\,\mathrm{m^2\,s^{-1}}$). This timescale is more than an order of magnitude smaller than the characteristic EDL charging time ($\sim10\,\mathrm{ns}$), implying that the velocity field adjusts quasi-instantaneously to the evolving electrostatic body force. Furthermore, owing to the nanometric channel dimensions and the resulting low electroosmotic velocities, the Reynolds number remains much smaller than unity, rendering convective inertia negligible. The hydrodynamics are therefore described by the incompressible Stokes equations throughout the transient EDL evolution. The momentum balance reduces to
\begin{align}
\dv{x}\!\left(\eta_v\dv{u}{x}\right)
+ E_l\rho_i = 0,
\qquad
x\in (-L+\lambda_s, L-\lambda_s)
\label{eq:EOF_dimensional}
\end{align}
where $\eta_v$ is the dynamic viscosity of the electrolyte and $u(x,t)$ is the velocity parallel to the charged surfaces. The no-slip boundary condition is imposed at the walls,
\begin{align}
u=0,
\qquad
x=\pm (L-\lambda_s).
\label{eq:NoSlip}
\end{align}

To account for the accompanying modification of momentum transport, the solvent viscosity is assumed to follow the experimentally validated viscoelectric relation \citep{lyklema1961interpretation,jin2022direct},
\begin{align}
\eta_v
=
\eta_0
\left(
1+f_vE^2
\right),
\end{align}
where $\eta_0$ is the zero-field viscosity and $f_v\!=\!1.02\!\times \! 10^{-15}\,\mathrm{m^2/V^2}$ is the viscoelectric coefficient. Together with the electrostatic model described in the previous subsection, this constitutes the Langevin--Bikerman-fitted transient (LBFT)--Stokes framework employed in the present study.

The electroosmotic response is characterized by the electroosmotic mobility $\mu_{eo}$, defined as the cross-sectionally averaged velocity in a half domain normalized by the applied lateral electric field,
\begin{align}
\mu_{eo}
=
\frac{1}{LE_l}
\int_{-L}^{0}
u(x)\,dx.
\label{eq:EOFmobility}
\end{align}
In the following sections, the LBFT--Stokes model is compared against the classical PNP--Stokes and MPNP--Stokes formulations, both of which employ a constant, field-independent viscosity $\eta_0$.

\subsection{Nondimensional parameters}

The transient electrostatic problem is nondimensionalized using the characteristic length $L$ and the electric double layer (EDL) charging time $\tau=\lambda_D L/D$, where $\lambda_D=
\sqrt{(\varepsilon_0\varepsilon_{r0})/(2n_0e_0^2\beta)}$ is the Debye length. This time scale characterizes the transient ion redistribution normal to the planar electrodes during EDL formation. The primary dimensionless variables are therefore
\begin{align}
\hat{x}=\frac{x}{L}, \qquad
\hat{t}=\frac{t}{\tau}, \qquad
\hat{\psi}=e_0\beta\psi,\qquad
\hat{n}_\pm=\frac{n_\pm}{2n_0},\qquad
\hat{\lambda}_s=\frac{\lambda_s}{L},
\end{align}
together with the dimensionless charge density $\hat{\rho}_i=\frac{\rho_i}{2n_0e_0}$ and dimensionless EDL thickness scale $\epsilon =\lambda_D/L$. The ion and solvent concentrations may also be expressed as molar concentrations, $c_0$ and $c_{0s}$, corresponding to the number concentrations  $n_0$ and $n_{0s}$, respectively. Hence, a relative bulk ion concentration parameter is defined as $\hat{c}_0= c_0/c_{0s}$.

Unlike the electrostatic problem, which is normalized using the characteristic EDL charging time, the electroosmotic flow parallel to the electrodes is normalized independently using the classical Helmholtz--Smoluchowski velocity scale, $u_{HS}\!=\!({\varepsilon_0\varepsilon_{r0}E_l})/({\eta_{0}e_0\beta})$, or equivalently the Helmholtz--Smoluchowski mobility, $\mu_{HS}\!=\!u_{HS}/E_l$, corresponding to a zeta potential of the order of the thermal voltage. The dimensionless velocity is $\hat{u}\!=\!u/u_{HS}$ and owing to the symmetry of the present configuration, the dimensionless electroosmotic mobility is therefore $\hat{\mu}_{eo} \!=\! \mu_{eo}/\mu_{HS} \!=\! \int_{0}^{1}\! \hat{u}\,d\hat{x}$.

Similarly, the electrohydrodynamic force density in the EDL region is normalized using the characteristic Maxwell stress scale
\begin{align}
\Pi_0 = \frac{\varepsilon_0\varepsilon_{r0}E_{EDL}^2}{\lambda_D},
\qquad
E_{EDL} = \frac{1}{e_0\beta\lambda_D},
\end{align}
which represents the characteristic electric field within the EDL. The corresponding dimensionless force density is defined as $\hat{\Pi}=\Pi/\Pi_0$. Finally, the strength of dipolar solvent polarization is characterized by the dimensionless parameter $\alpha \! = \! \beta p_0 E_0$. For weak dipolar alignment ($\alpha\ll1$), the Langevin--Bikerman formulation recovers the bulk relative permittivity, $\varepsilon_r \approx \varepsilon_{r0}$, corresponding to the modified Poisson--Nernst--Planck (MPNP) limit described by \citet{kilic2007bsteric}.

\subsection{Numerical Implementation}
Solving the time-dependent LBFT equations under a sudden voltage step is implemented in the 1D model in COMSOL Multiphysics v6.2, leveraging its state-of-the-art finite element method (FEM) framework. The Nernst--Planck equations are discretized using quadratic (P2) elements, while the Poisson equation uses cubic (P3) elements to capture sharp spatial variations in the electrostatic potential and polarization fields. A highly refined nonuniform mesh is employed: the near-wall region $x\in\pm(L, L-10\lambda_D)$ is discretized with 1500 equally spaced elements, and the remaining bulk domain with elements of size $\lambda_D/25$, ensuring grid-independent results.

Time integration is performed using the generalized-$\alpha$ method, a second-order implicit scheme designed for stiff, transient PDEs. The maximum time step is restricted to $ \tau\times10^{-3}$, with a relative solver tolerance of $2\%$. To maintain stability without sacrificing accuracy during fast charging phases, the high-frequency dissipation factor in the $\alpha$-method is tuned between 0.5--0.75, selectively damping numerical noise without suppressing true dynamics.

At each time step, the nonlinear system is solved using Newton’s method, with up to 8 iterations and an adaptive line search strategy. The “Highly nonlinear Newton” setting to enhance robustness via adaptive step-size damping and automatic backtracking. In particular, a recovery damping factor of 0.2--0.75 is employed when residuals temporarily increase, allowing safe step-size reduction and recovery from divergence. A minimum damping factor of 0.001 ensures conservative updates when strong nonlinearity is encountered. If convergence stalls, the Newton step size is reduced by a factor of 10 before reattempting. The resulting linearized system is solved at each Newton step using the MUMPS (Multifrontal Massively Parallel Sparse) direct solver, which performs robust LU factorization optimized for large, sparse, and ill-conditioned matrices. 


Simulations at five combinations of parameters relevant for modern voltage-gated applications such as nanofluidic logic and biosensing, listed in table \ref{tab:1Cases}, are conducted using the three models, LBFT-Stokes, MPNP-Stokes, and PNP-Stokes. These cases of parameters and the anticipated time scales of the EDL dynamics and transient EOF dynamics are of the order of $10$ ns, relevant to voltage-gated nanofluidic applications\citep{he2011controlling}. Case I indicates a domain of length $25$ nm, with molecular dipole moment of the solvent close to that of the effective dipole moment of water\citep{abrashkin2007dipolar}. Case II demonstrates a reduced channel width ($L = 10$ nm), case III has a slightly weaker dipole moment yet relevant to aqueous electrolytes ($p_0 = 4.75$ D), case IV has a thick EDL due to lower ion concentration ($c_0=0.1$ mM), and case V has a lower applied voltage ($V_0=20$ mV). Thus, comparisons of results from these cases serve for isolating different parametric influences.
\begin{table}
  \centering
  \caption{Simulation parameters for the four representative cases considered in this study, selected to reflect relevant application conditions. The temperature is fixed at 290 K for all cases with the characteristic ion diffusivity of $10^{-9}$ m$^2$/s. \\}
  \begin{tabular}{@{}cllllllllcl@{}}
    \toprule
    Case & $L$ (nm) & $c_0$ (mM) & $p_0$ (D) & $V_0$ (V) & $\tau$ (ns)& $\epsilon$ & $\alpha$ & $\hat{c}_0$ & $\hat{V}_0$ & Remarks  \\
    \midrule
    I   & 25 & 100  & 4.95 & 0.25 & 25 & 0.039 & 0.0041 & 3.6 & 10 & wider domain, lower $\alpha$ \\ 
    II  & 10 & 100  & 4.95 & 0.10 & 10 & 0.099 & 0.0103 & 3.6 & 4 &  parametric case for comparison \\
    III & 10 & 100  & 4.75 & 0.10 & 10 & 0.095 & 0.0099 & 3.6 & 4 & lower solvent dipole moment \\
    IV  & 10 & 0.1  & 4.95 & 0.10 & 300 & 3.144 & 0.0103 & 0.0036 & 4 & lower concentration \\
    V  & 10 & 100  & 4.95 & 0.01 & 10 & 0.099 & 0.0103 & 3.6 & 0.4 & lower voltage \\
    \bottomrule
  \end{tabular} \label{tab:1Cases}
\end{table}

\section{\label{sec:results}Results and Analysis}
\subsection{Transient EDL Dynamics}

\subsubsection{Space Charge Distribution}
\begin{figure}
\centering
\begin{tikzpicture}
  \sbox0{\includegraphics[width=2.0in]{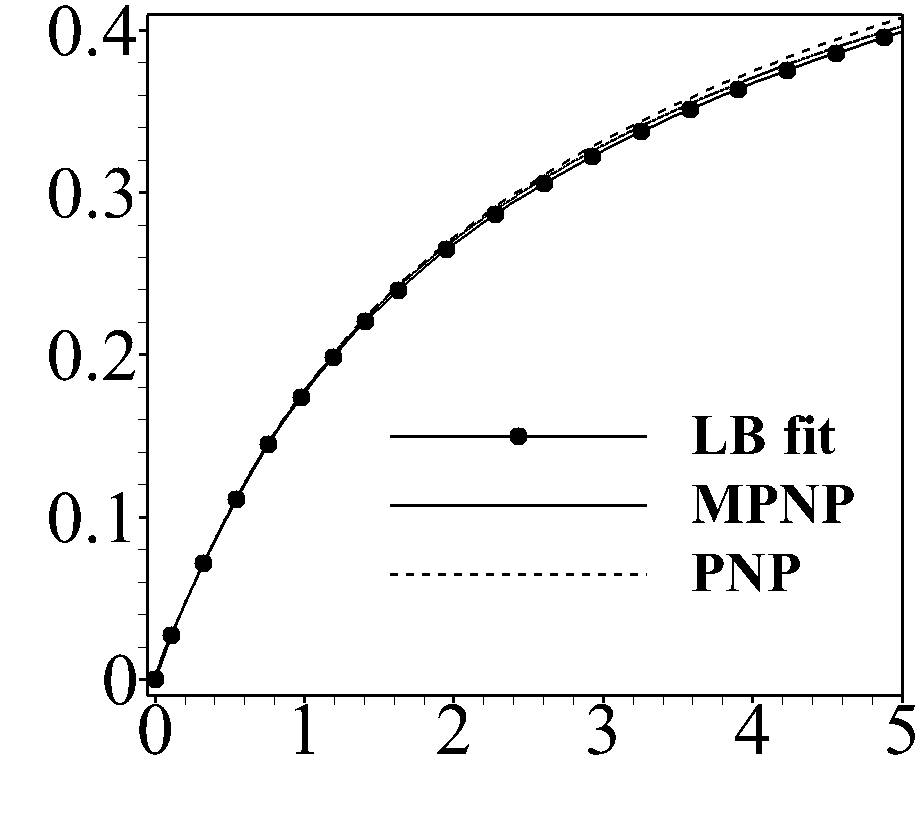} } 
  \node[above right,inner sep=0pt] at (0,0)  {\usebox{0}};
  \node[black] at (0.58\wd0,0.02\ht0) {$\hat{t}$};
    \node[black,rotate=90] at (0.0\wd0,0.55\ht0) {$\hat{q}$};
    \node[black] at (0.45\wd0,0.9\ht0) {$\hat{\tau}_q=2.65\pm0.04$};
  \node[black] at (0.0\wd0,1.0\ht0) {(a) };
\end{tikzpicture}
\begin{tikzpicture}
  \sbox0{\includegraphics[width=2.0in]{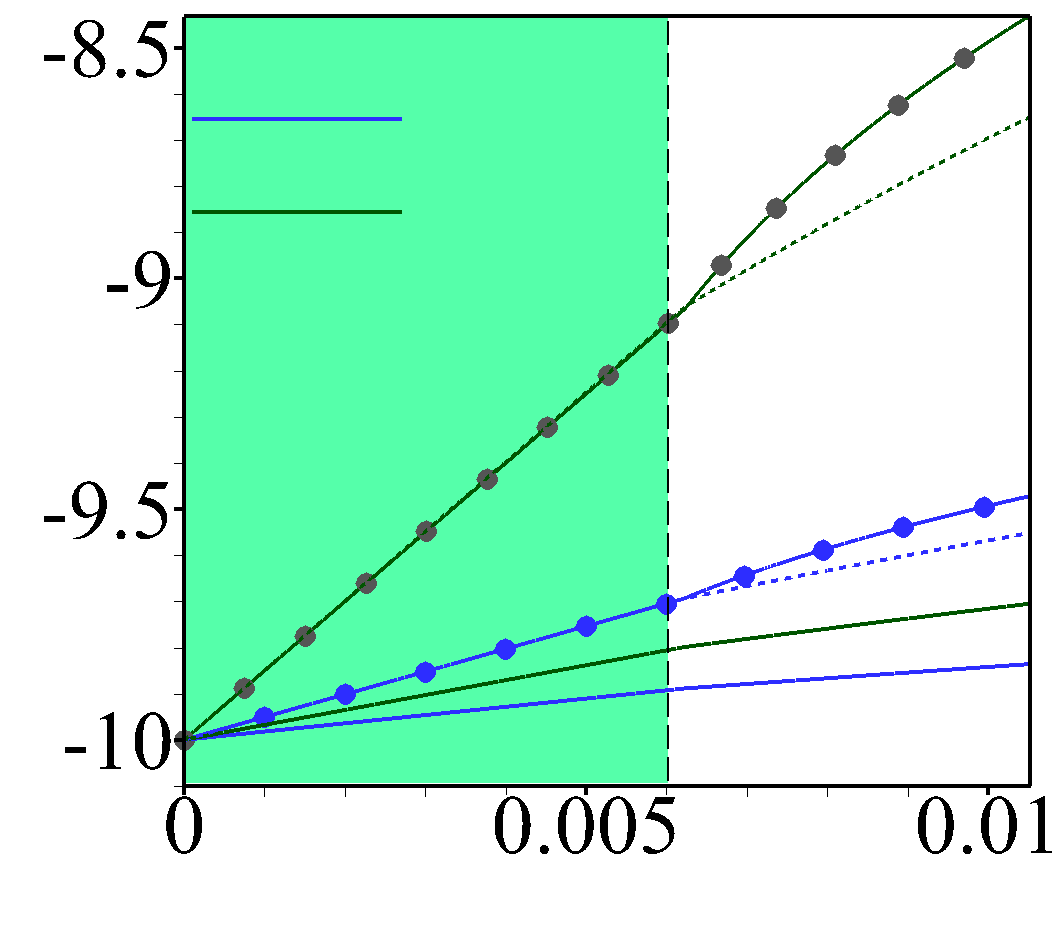} } 
  \node[above right,inner sep=0pt] at (0,0)  {\usebox{0}};
  \node[black] at (0.58\wd0,0.02\ht0) {$\hat{x}+1$};
  \node[black,rotate=90] at (0.02\wd0,0.58\ht0) {$\hat{\psi}$};
  \node[blue] at (0.48\wd0,0.89\ht0) {$\hat{t}=0.1$ };
  \node[red] at (0.48\wd0,0.79\ht0) {$\hat{t}=0.5$ };
  \node[black] at (-0.01\wd0,1.0\ht0) {(b) };
\end{tikzpicture}
\caption{(a) Temporal evolution of the total EDL charge with fitted relaxation time scale $\tau_q$ predicted by the different models for Case I. (b) Spatial variation of the electric potential as a function of distance from the electrode surface for Case I, shown at two representative times. The shaded region $\hat{x\,}<\lambda_s\!=\!0.006$ corresponds to the Stern layer. (Color online.)}
\label{fig:2}
\end{figure}
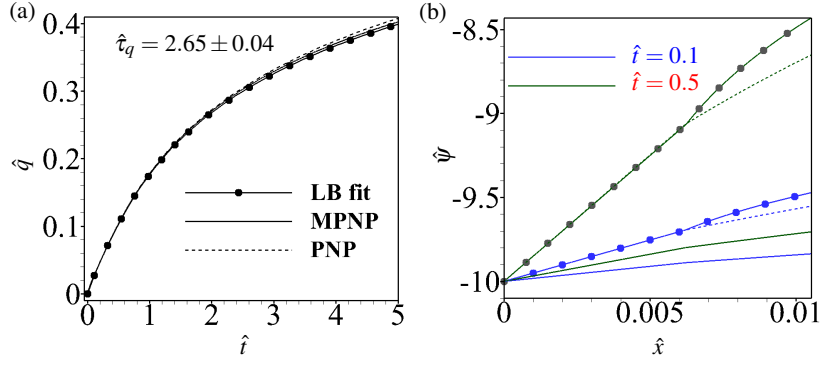

The nonuniform permittivity arising from dipolar solvent response modifies the charging dynamics of the electric double layer (EDL). The total diffuse charge density accumulated near the surface, beyond the Stern layer, is quantified as $\hat{q}=\int_{-1+\lambda_s/L}^{-1+10\lambda_D/L} \hat{\rho}_i\, d\hat{x}$. 
Since the net charge density decays exponentially over the Debye length $\lambda_D$, the overwhelming majority of the excess charge resides within a few Debye lengths of the surface. Hence, integrating over a distance of $10\lambda_D$ therefore captures essentially the entire EDL charge while excluding the electroneutral bulk, providing a consistent and physically meaningful basis for comparing the different continuum models.
For Case~I, the value of $\hat{q}$ predicted by the LBFT model closely follows those obtained from the MPNP and classical PNP formulations, as shown in Fig.~\ref{fig:2}(a). Any noticeable deviations emerge only at later times, where the LBFT estimate is slightly higher than the MPNP prediction. The exponentially fitted dimensionless charging time scales ($\hat{\tau}_q$) remain nearly identical across all models, consistent with steric-modified PNP behavior reported by \citet{kilic2007bsteric}. These results indicate that dipolar solvent corrections primarily affect the magnitude of the accumulated charge while leaving the global charging time constant essentially unchanged.

The Fig.~\ref{fig:2}(b) shows the electric potential distribution in the vicinity of the electrode, including the Stern layer, at two representative instances, $\hat{t}=0.1,\hat{t}=0.5$ of transient EDL charging for Case~I. As the diffuse charge develops, the electric field at the Stern--diffuse interface evolves continuously to satisfy the coupled electrostatic response of the Stern and diffuse layers under the externally prescribed electrode potentials. Consequently, the total voltage drop is redistributed dynamically between these two regions during charging. The electric potential at the Stern--electrolyte interface ($\hat{x}=0.006$) predicted by the LBFT model differs systematically from the MPNP and PNP results owing to the field-dependent dielectric response arising from dipolar solvent polarization.

At the Stern--diffuse interface, both the electric potential and the normal electric displacement remain continuous, whereas the electric field, represented by the negative slope of the potential profile in Fig. \ref{fig:2}(b), changes according to the permittivity on either side of the interface. For the PNP and MPNP models, the permittivity within each region is prescribed to two different constant values, so the field ratio across the interface is determined solely by the corresponding permittivity ratio. The MPNP model further modifies the potential distribution (and hence the field values) through steric-induced changes in the diffuse-space-charge profile. In contrast, the LBFT model allows the diffuse-layer permittivity to evolve with the local electric field, leading to a continuously varying electric-field ratio across the interface throughout the transient charging process. This behavior extends the steady-state observations reported previously by \citet{srinivasula2025dipolar} to the transient regime.

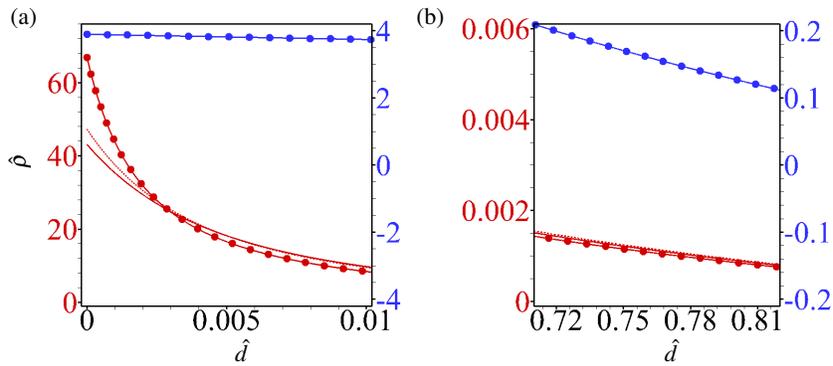
\begin{figure}
\centering
\begin{tikzpicture}
  \sbox0{\includegraphics[width=2.0in]{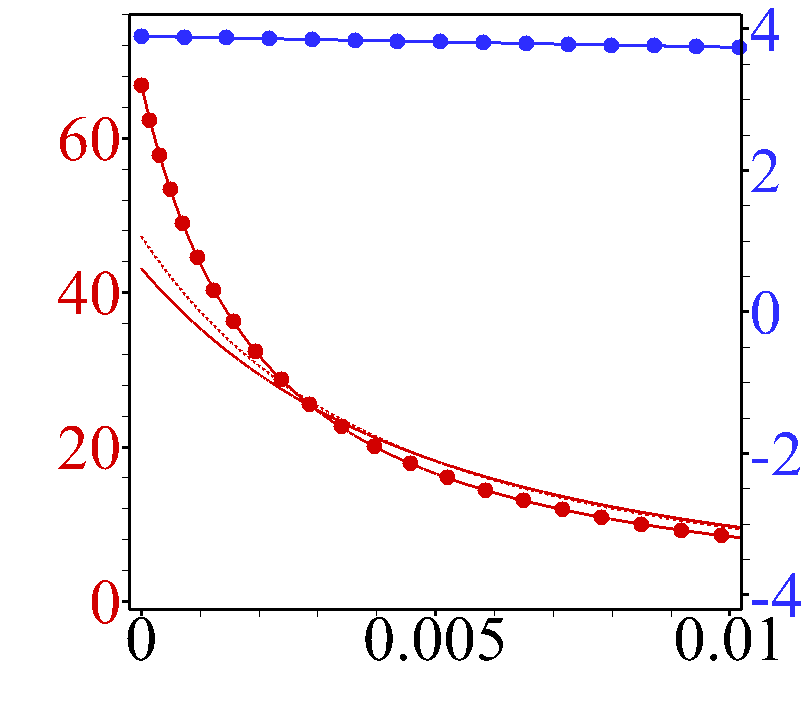} } 
  \node[above right,inner sep=0pt] at (0,0)  {\usebox{0}};
  \node[black] at (0.58\wd0,0.02\ht0) {$\hat{x}-\hat \lambda_s+1$};
  \node[black,rotate=90] at (0.00\wd0,0.58\ht0) {$\hat{\rho}_i$ };
  \node[black] at (0.01\wd0,1.0\ht0) {(a) };
\end{tikzpicture}
\begin{tikzpicture}
  \sbox0{\includegraphics[width=2.0in]{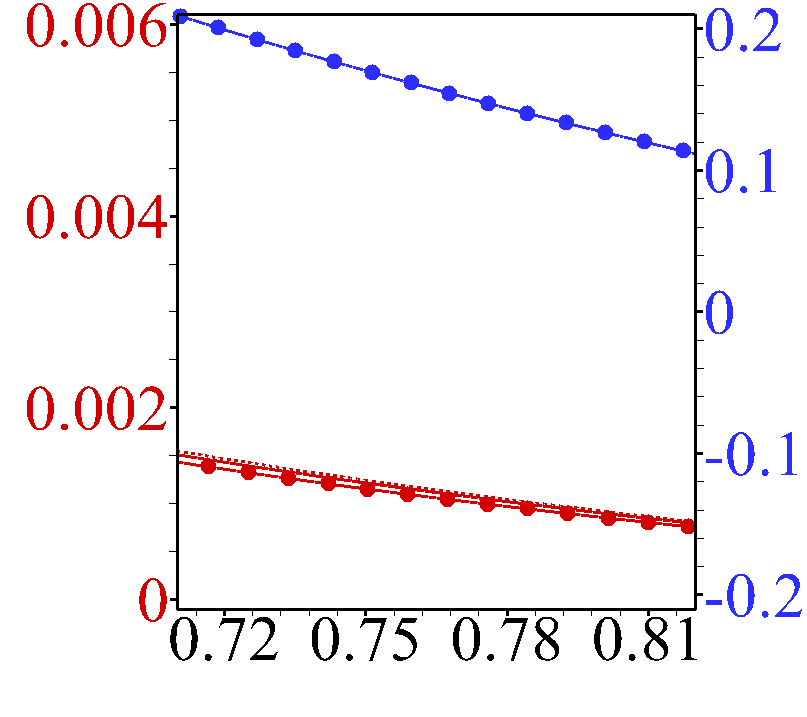} } 
  \node[above right,inner sep=0pt] at (0,0)  {\usebox{0}};
  \node[black] at (0.58\wd0,0.02\ht0) {$\hat{x}-\hat \lambda_s+1$};
  \node[black] at (-0.05\wd0,1.0\ht0) {(b) };
\end{tikzpicture}

\caption{(a, b) Distributions of charge density from different models, at $\hat{t} = 2.7$ for case I (red) \& IV (blue) near and away from the surface, respectively. See Figure \ref{fig:2}(a) for the legend. (Color online.)}
\label{fig:3_ChargeDistrib}
\end{figure}

The instantaneous charge structure is further examined in Fig.~\ref{fig:3_ChargeDistrib} at an intermediate time $\hat{t}=2.7$ for Cases~I and~IV, both near and away from the Stern boundary. Owing to antisymmetry about the channel mid-plane ($\hat{x}=0$), only one half of the domain is shown. The charge density near the negatively charged electrode is plotted as a function of the distance from the Stern--EDL interface located at $\hat{x}=-1+\hat\lambda_s$.
In the near-surface region [Fig.~\ref{fig:3_ChargeDistrib}(a)], the LBFT model predicts a higher ionic charge density for Case~I. This increase arises from reduced dielectric screening under strong electric fields, which enhances charge accumulation close to the interface. In contrast, Case~IV corresponds to a dilute electrolyte with a comparatively thick EDL, where the Debye length exceeds the characteristic domain scale due to the smaller value of $\alpha$ (representative of typical aqueous electrolytes). Under these conditions, the LBFT predictions become nearly indistinguishable from those of the MPNP and PNP models.

Farther from the surface [Fig.~\ref{fig:3_ChargeDistrib}(b)], the electric field is largely screened by the diffuse layer. To maintain global charge balance, the LBFT model yields a slightly lower charge density in the outer region relative to the MPNP and PNP predictions. Although these spatial redistributions are subtle, the present results indicate their significant influence on the transient behavior of the system. 

\begin{figure}
\centering
\begin{tikzpicture}
  \sbox0{\includegraphics[width=2.0in]{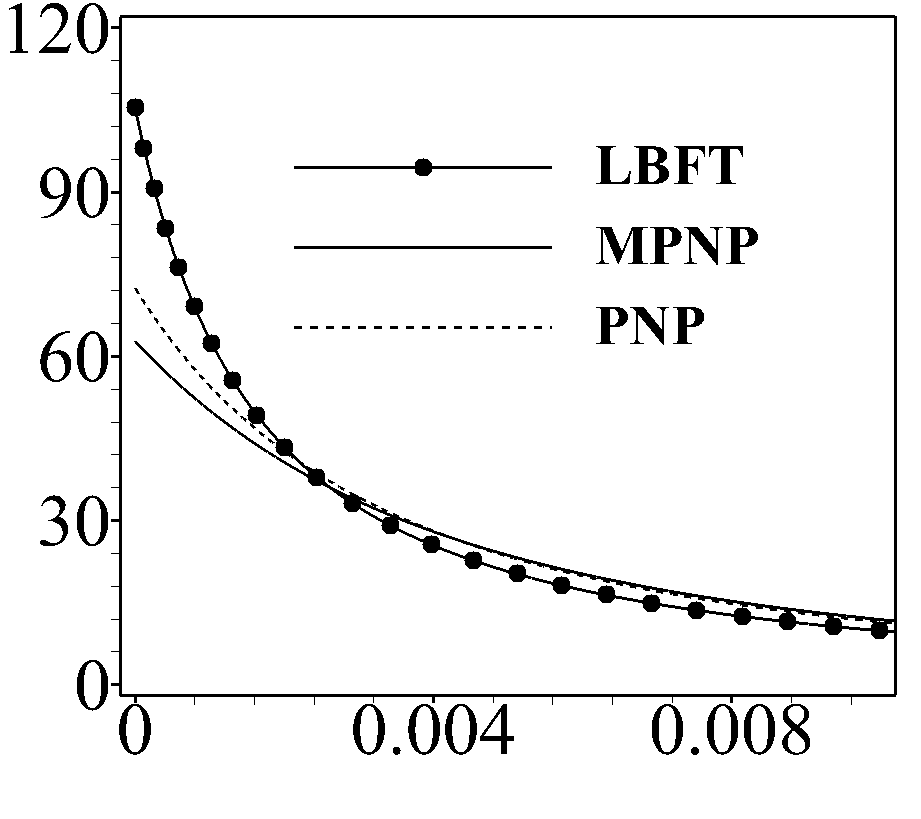} } 
  \node[above right,inner sep=0pt] at (0,0)  {\usebox{0}};
  \node[black] at (0.58\wd0,0.02\ht0) {$\hat{x}-\hat \lambda_s+1$};
  \node[black,rotate=90] at (-0.02\wd0,0.58\ht0) {$\hat{\rho}_i$};
  \node[black] at (-0.05\wd0,1.0\ht0) {(a) };
\end{tikzpicture}
\begin{tikzpicture}
  \sbox0{\includegraphics[width=2.0in]{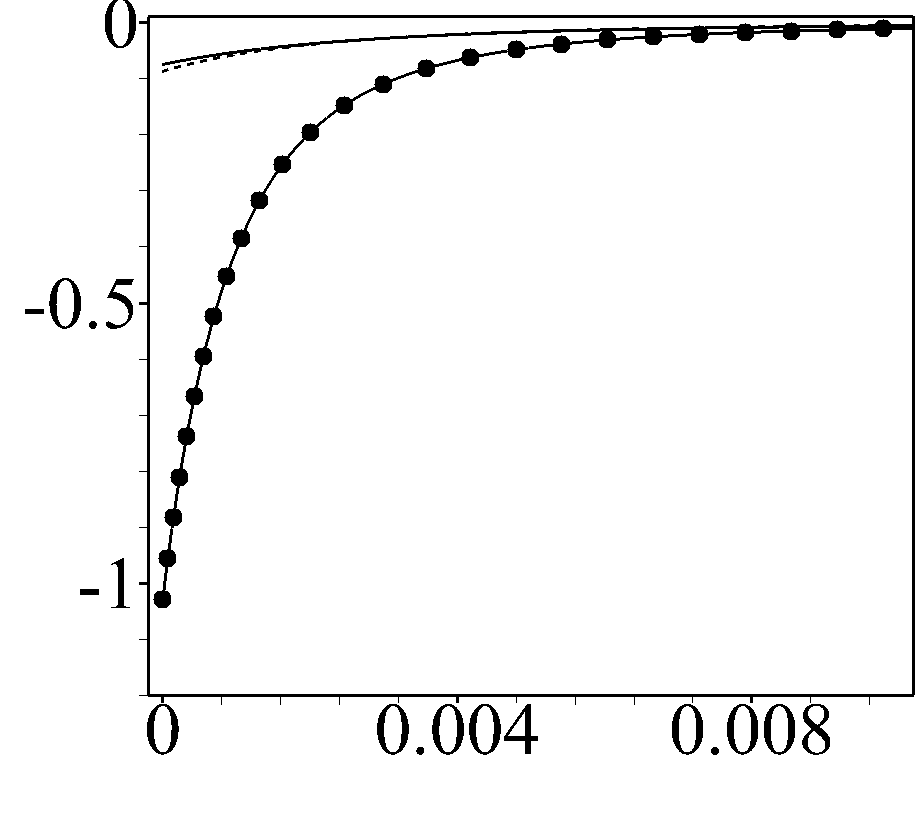} } 
  \node[above right,inner sep=0pt] at (0,0)  {\usebox{0}};
  \node[black] at (0.58\wd0,0.02\ht0) {$\hat{x}-\hat \lambda_s+1$};
  \node[black,rotate=90] at (-0.03\wd0,0.6\ht0) {$\hat{\Pi}$};
  \node[black] at (0.0\wd0,1.0\ht0) {(b) };
\end{tikzpicture}
\caption{(a) The scaled space charge density and (b) normalized EHD force density profiles at $\hat{t} = 5$ for case I. (Color online.)}
\label{fig:4_Eforce}
\end{figure}
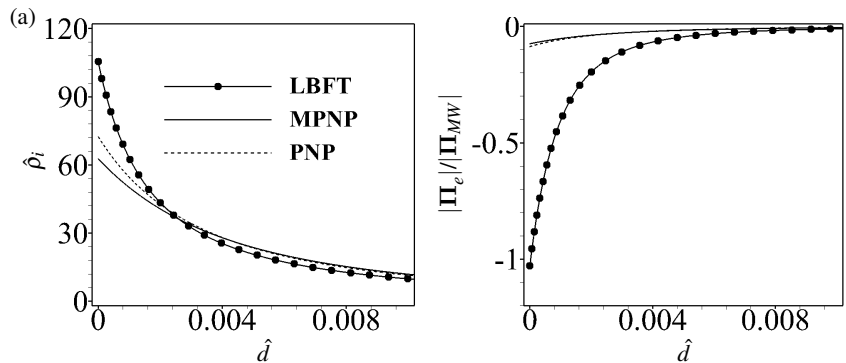

\subsubsection{Normal Electrohydrodynamic Force Density within EDL}

The space--charge distribution and the resulting electrohydrodynamic (EHD) force density are shown in Fig.~\ref{fig:4_Eforce}(a,b) at $\hat{t}=5.0$ for Case~I. The LBFT model predicts a substantially enhanced near-surface Maxwell force density, reaching values up to an order of magnitude larger than those obtained from the classical PNP and MPNP formulations. This relative amplification is more pronounced than the corresponding increase in ionic charge density and arises primarily from the dipolar solvent contribution under dielectric saturation, which strengthens the local electric field gradients adjacent to the electrode.
In the present configuration, the normal electrohydrodynamic force density, $\Pi_x$ [Fig.~\ref{fig:4_Eforce}(b)], does not drive fluid motion. Instead, it contributes to the measurable surface force acting on the electrodes, providing an additional observable for quantifying the dipolar solvent corrections identified in this study.

\subsection{Transient Electrohydrodynamic Response}
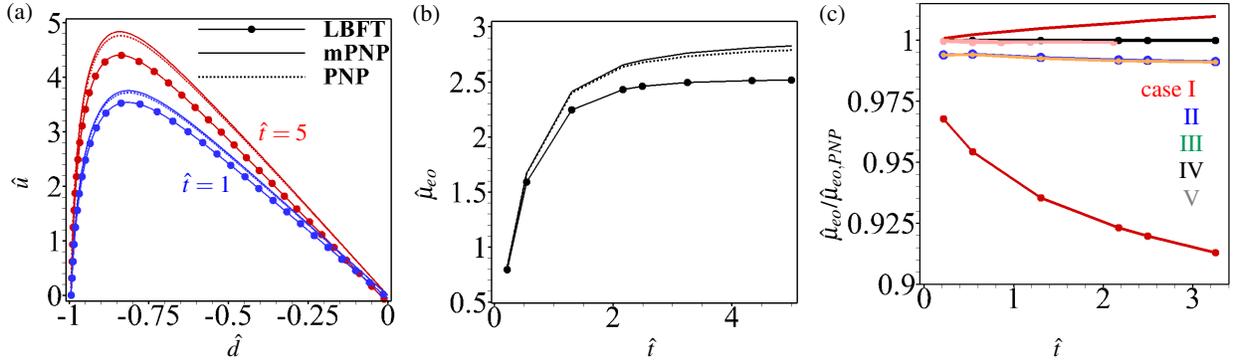
\begin{figure*}
\centering
\begin{tikzpicture}
  \sbox0{\includegraphics[width=2.0in]{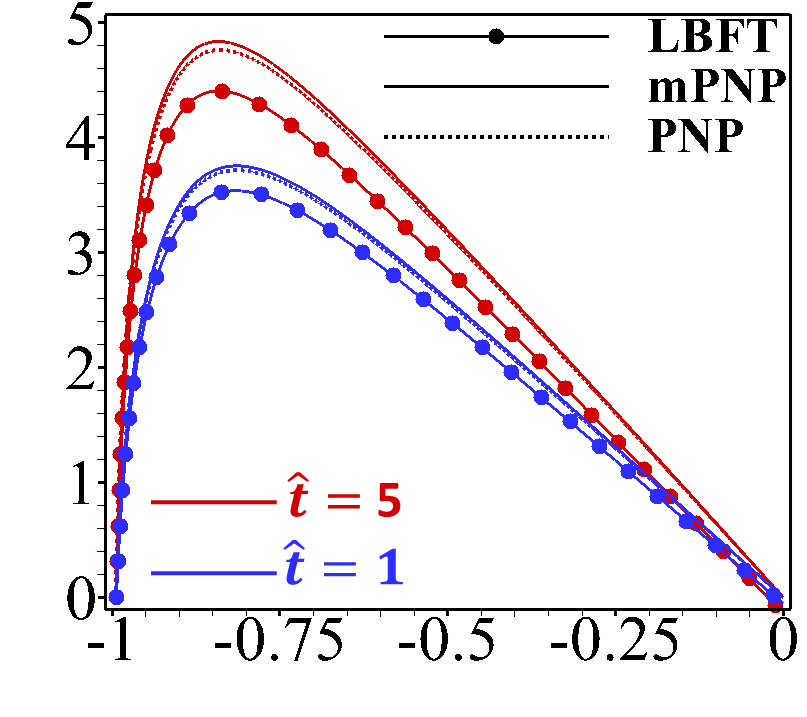} }
  \node[above right,inner sep=0pt] at (0,0)  {\usebox{0}};
  \node[black] at (0.57\wd0,0.02\ht0) {$\hat{x}-\hat \lambda_s+1$};
  \node[black,rotate=90] at (0.01\wd0,0.5 \ht0) {$\hat{u}$};
  \node[black] at (0.01\wd0,1.0\ht0) {(a)};
\end{tikzpicture}
\begin{tikzpicture}
  \sbox0{\includegraphics[width=2.0in]{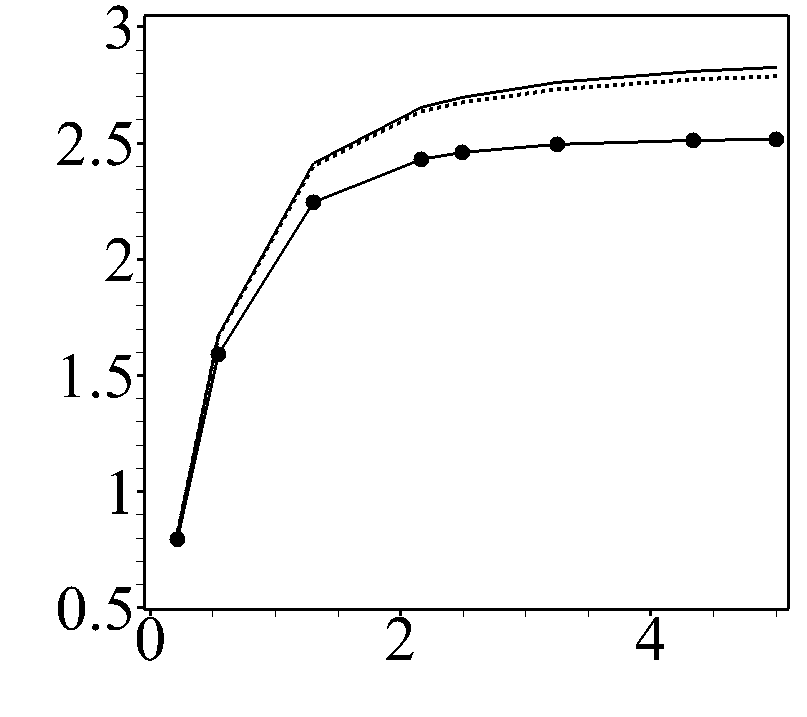} }
  \node[above right,inner sep=0pt] at (0,0)  {\usebox{0}};
  \node[black] at (0.6\wd0,0.02\ht0) {$\hat{t}$};
  \node[black,rotate=90] at (0.01\wd0,0.5 \ht0) {$\hat{\mu}_{eo}$};
  \node[black] at (0.01\wd0,1.0\ht0) {(b) };
\end{tikzpicture}
\begin{tikzpicture}
  \sbox0{\includegraphics[width=2.0in]{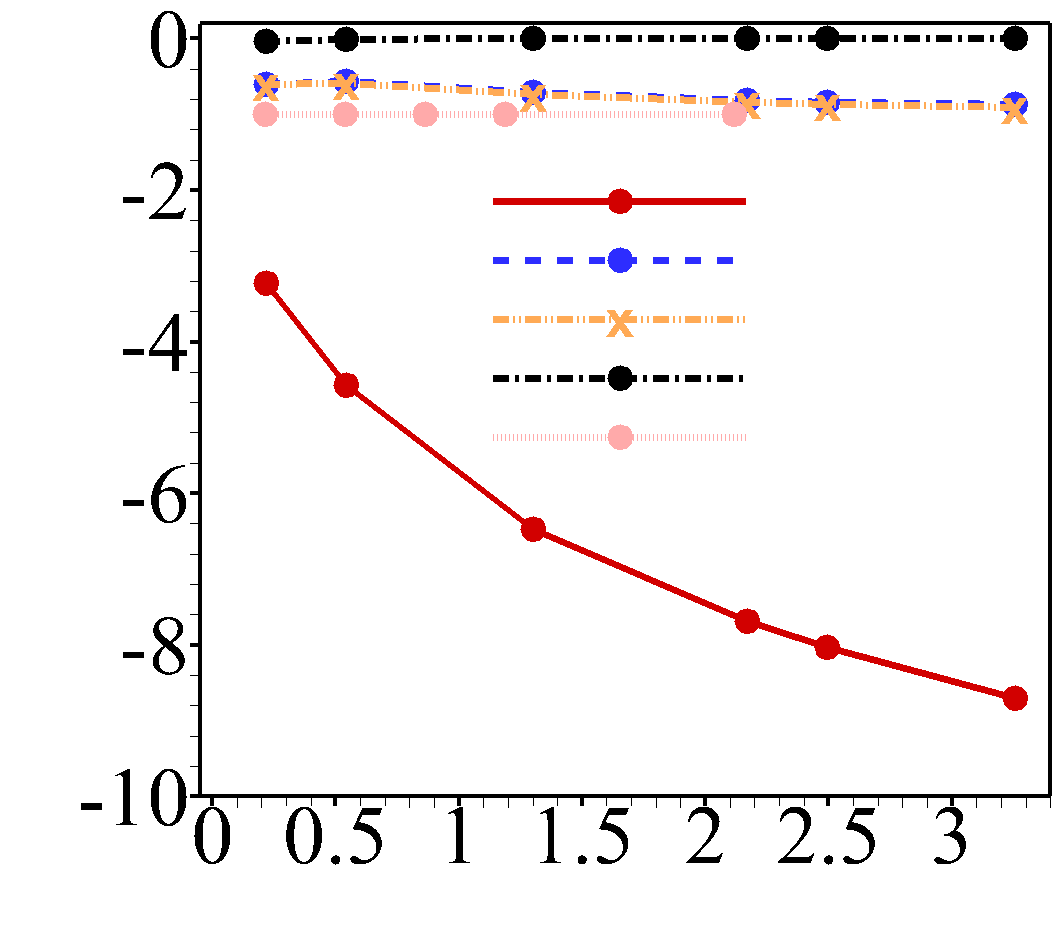} }
  \node[above right,inner sep=7pt] at (0,0)  {\usebox{0}};
  \node[black] at (0.6\wd0,0.02\ht0) {$\hat{t}$};
  \node[black,rotate=90] at (0.01\wd0,0.5 \ht0) {$\hat{\mu}_{eo}$/$  \hat{\mu}_{eo,PNP}$};
  \node[black] at (0.01\wd0,1.0\ht0) {(c) };
  \node[red,rectangle, inner sep=0pt] at (0.89\wd0,0.85\ht0) {case I};
  \node[blue] at (0.9\wd0,0.79\ht0) {case II};
  \node[ForestGreen, rectangle, inner sep=0pt] at (0.91\wd0,0.72\ht0) {case III };
  \node[black, rectangle, inner sep=0pt] at (0.92\wd0,0.66\ht0) {case IV};
  \node[gray, rectangle, inner sep=0pt] at (0.91\wd0,0.59\ht0) {case V};
\end{tikzpicture}
 \caption{Transient EOF with dielectric saturation and no viscoelectric effect of LBFT model. (a) Spatio-temporal profiles of EOF velocity from different models for case I and corresponding (b) EOF mobility evolutions with time. (c) Evolution of correction factors for EOF mobility for LBFT, MPNP models relative to PNP for the different cases shown in different colors. (Color online.)}
\label{fig:5_EOF_NoVE}
\end{figure*}

\subsubsection{Dielectric Saturation Effect on Transient Electroosmotic Flow}
The velocity profiles predicted by the different electrostatic models, excluding the viscoelectric correction ($f_v=0$) in the Stokes equation, are shown in Fig.~\ref{fig:5_EOF_NoVE}(a) at two representative instants during the spatiotemporal evolution of electroosmotic flow (EOF). Under the no-slip boundary condition at the wall, the modest reduction in charge density away from the surface outweighs the pronounced near-surface increase observed in Fig.~\ref{fig:3_ChargeDistrib}. Consequently, the LBFT model predicts a slightly lower bulk EOF velocity compared with the classical PNP and modified PNP (MPNP) formulations. This trend is consistent with the steady-state comparison reported in \citet[Fig.~4]{srinivasula2025dipolar}, where the Langevin--Bikerman model yields a reduced EOF velocity relative to the Poisson--Boltzmann limit, particularly in the weak dipolar regime ($\alpha \ll 1$). The influence of dielectric saturation on the transient velocity profiles becomes progressively more pronounced as the system evolves toward steady state.

The corresponding transient evolution of the normalized electroosmotic mobility, $\hat{\mu}_{eo}$, is shown in Fig.~\ref{fig:5_EOF_NoVE}(b). All models exhibit a monotonic increase toward their respective steady-state values. The MPNP results remain close to PNP, indicating that ion steric effects introduced only minor quantitative changes to the EOF mobility. However, the LBFT prediction consistently remains lower, reflecting the cumulative influence of dielectric saturation on the effective electrohydrodynamic driving force within the diffuse layer.

Figure~\ref{fig:5_EOF_NoVE}(c) presents the temporal variation of the mobility ratios predicted by the LBFT model relative to the classical PNP formulation across parametric Cases~I--V. The dipolar corrections incorporated in the LBFT model are more pronounced, particularly for Case~IV, which corresponds to a comparatively thick EDL. In this regime, dipolar modifications to the diffuse charge extend farther into the bulk, leading to a larger reduction in electroosmotic mobility as the system approaches steady state, although the dielectric saturation itself is weaker due to the lower near-surface charge density and electric field.

\subsubsection{Viscoelectric Effect on Transient Electroosmotic Flow}
\begin{figure}
\centering
\begin{tikzpicture}
  \sbox0{\includegraphics[width=2.0in]{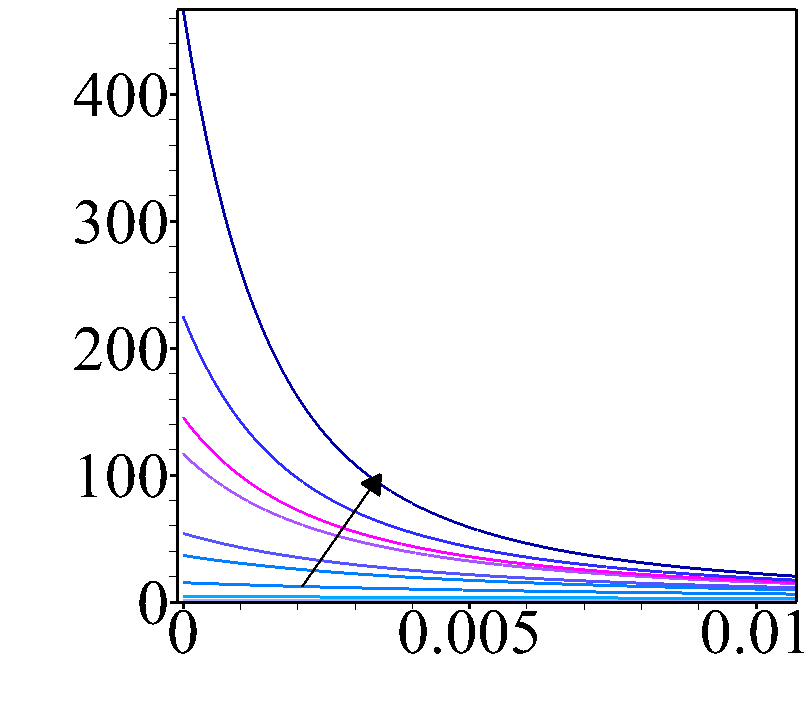} } 
  \node[above right,inner sep=0pt] at (0,0)  {\usebox{0}};
  \node[black] at (0.58\wd0,0.02\ht0) {$\hat{x}-\hat \lambda_s+1$};
  \node[black,rotate=90] at (-0.02\wd0,0.58\ht0) {$\hat{\eta}_v$};
  \node[black] at (0.0\wd0,1.0\ht0) {(a) };
  \node[black] at (0.5\wd0,0.4 \ht0) {$\hat{t}$};
\end{tikzpicture}
\begin{tikzpicture}
  \sbox0{\includegraphics[width=2.0in]{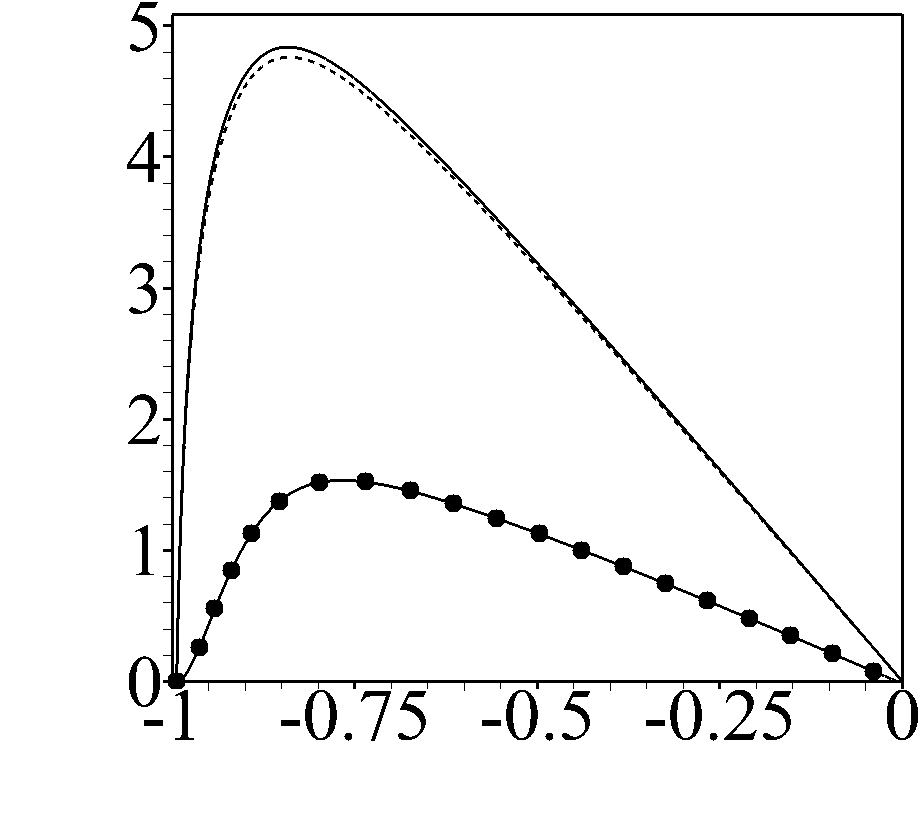} } 
  \node[above right,inner sep=0pt] at (0,0)  {\usebox{0}};
  \node[black] at (0.58\wd0,0.02\ht0) {$\hat{x}-\hat \lambda_s+1$};
  \node[black,rotate=90] at (-0.02\wd0,0.58\ht0) {$\hat{u}$};
  \node[black] at (0.0\wd0,1.0\ht0) {(b) };
\end{tikzpicture}
\begin{tikzpicture}
  \sbox0{\includegraphics[width=2.0in]{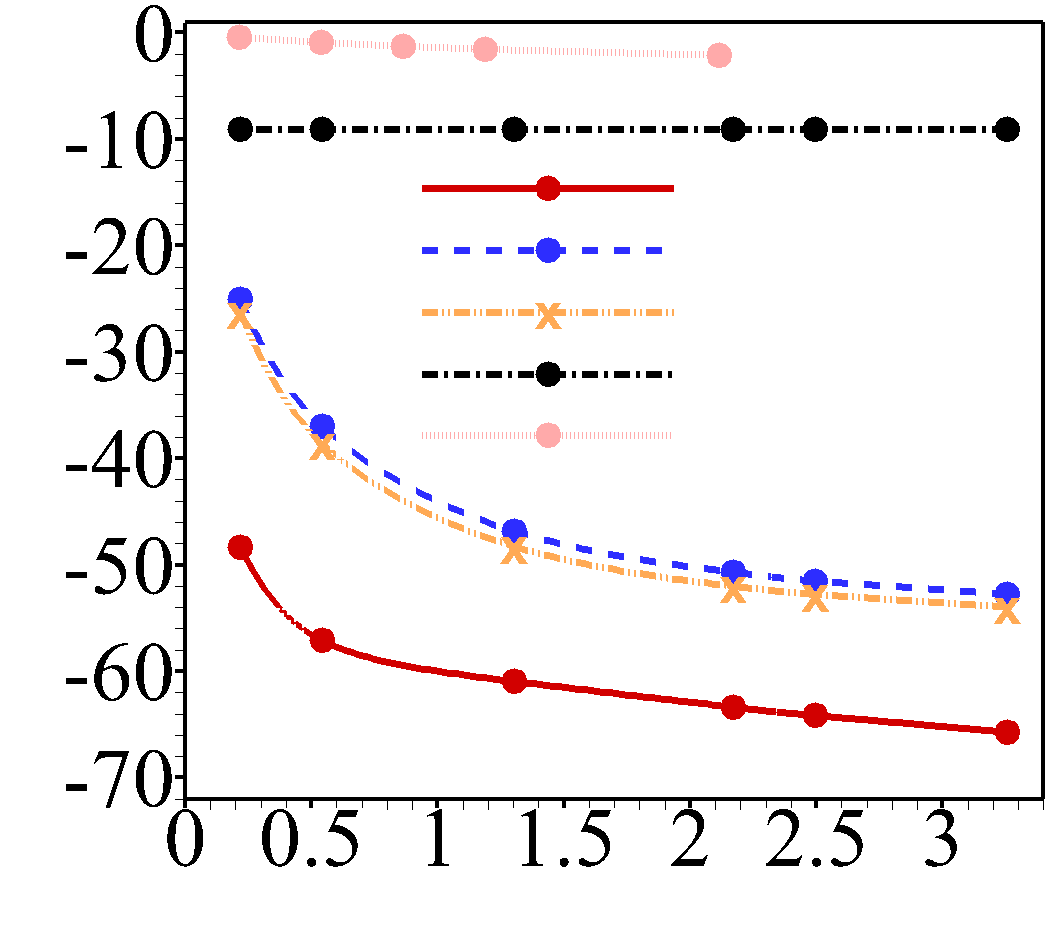} } 
  \node[above right,inner sep=0pt] at (0,0)  {\usebox{0}};
  \node[black] at (0.58\wd0,0.02\ht0) {$\hat{t}$};
  \node[black,rotate=90] at (-0.02\wd0,0.58\ht0) {$\hat{\mu}_{eo}$/$\hat{\mu}_{eo,PNP}$};
  \node[black] at (0.0\wd0,1.0\ht0) {(c) };
  \node[red,rectangle, inner sep=0pt] at (0.75\wd0,0.8\ht0) {case I};
  \node[blue] at (0.76\wd0,0.74\ht0) {case II};
  \node[ForestGreen, rectangle, inner sep=0pt] at (0.77\wd0,0.67\ht0) {case III };
  \node[black, rectangle, inner sep=0pt] at (0.78\wd0,0.61\ht0) {case IV};
  \node[gray, rectangle, inner sep=0pt] at (0.77\wd0,0.54\ht0) {case V};
\end{tikzpicture}
\caption{Transient EOF with dielectric saturation and viscoelectric effects of LBFT model compared to traditional PNP/MPNP- Stokes models. (a) Spatio-temporal evolution of viscosity in LBFT model, (b) velocity profile estimates from different models at $\hat{t}=5$ for case I. (c) EOF mobility correction factors of LBFT with viscoelectric effects compared to PNP model, at $\hat{t}=5$. (Legends same as figure \ref{fig:5_EOF_NoVE}.) (Color online.)}
\label{fig:6_EOF_u_VE}
\end{figure}
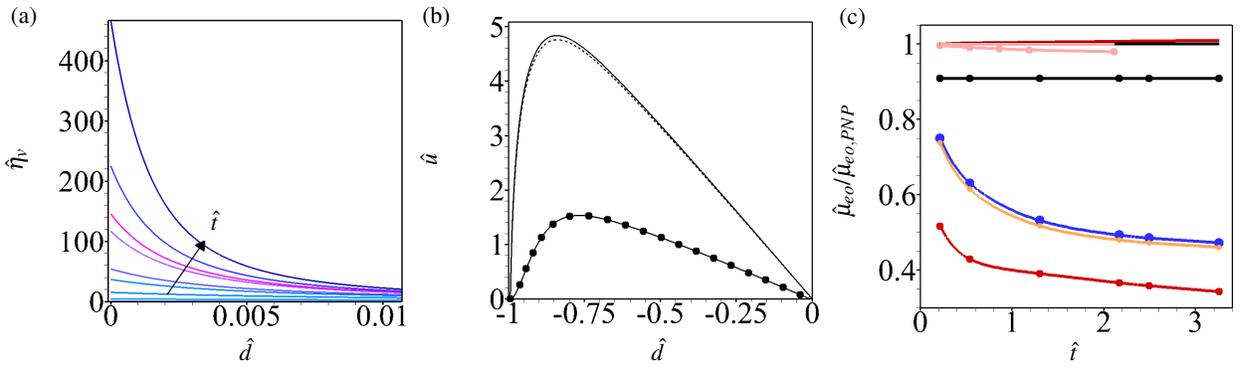
Figure~\ref{fig:6_EOF_u_VE} presents the electroosmotic flow (EOF) predicted by the LBFT model with viscoelectric effects included for an aqueous-like dipolar solvent, using the coefficient $f_v = 1.02~\mathrm{m^2/V^2}$ \citep{jin2022direct, lyklema1961interpretation, hsu2016electrokinetics}. The spatiotemporal evolution of the viscosity field is shown in Fig.~\ref{fig:6_EOF_u_VE}(a). A pronounced viscosity enhancement develops within the electric double layer (EDL), where the electric field is strongest and dipolar alignment is most significant during the transient EDL evolution. Outside the EDL, the viscosity rapidly relaxes toward its bulk value as the electric field weakens, indicating that the viscoelectric correction remains strongly localized near the interface.

The corresponding EOF velocity profiles at $\hat{t}=5$ predicted by the different electrostatic models are shown in Fig.~\ref{fig:6_EOF_u_VE}(b). The increased near-wall viscosity in the LBFT formulation substantially reduces the peak velocity while preserving the qualitative ordering among the models. In addition to lowering the overall flow magnitude, the viscoelectric contribution modifies the transient electroosmotic response through the combined action of dielectric saturation and field-dependent viscous resistance, both arising from the dipolar nature of the solvent.

This combined influence is further illustrated in Fig.~\ref{fig:6_EOF_u_VE}(c), which shows the temporal evolution of the electroosmotic mobility correction ratios. Cases~I and~IV, characterized by comparatively weaker nominal electric fields, exhibit moderate corrections or 2--10\%, whereas the remaining cases display increasingly pronounced deviations growing with time up to 65\%, as the system approaches steady state.

Overall, these results demonstrate that the coupled dielectric and viscoelectric responses of a dipolar solvent can significantly modify transient nanoscale electroosmotic transport. Neglecting either mechanism may therefore lead to systematic overprediction of electroosmotic velocities in strongly driven nanofluidic regimes.

\subsection{Very-high-frequency AC electrokinetic flow}

\begin{figure}
\centering
\begin{tikzpicture}
  \sbox0{\includegraphics[width=2.0in]{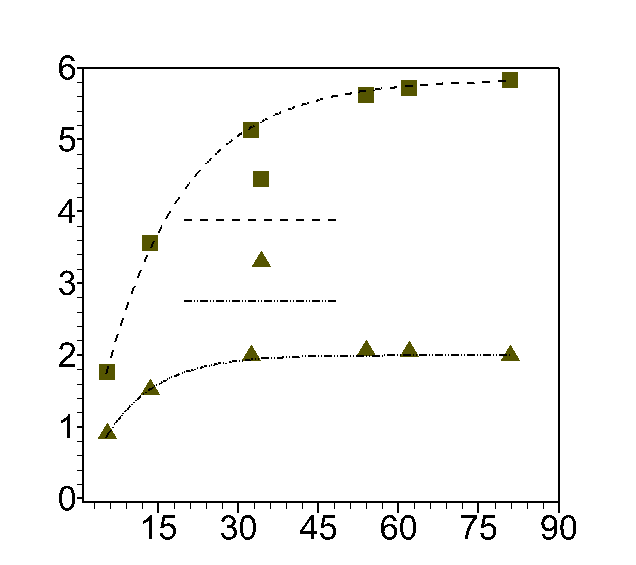} } 
  \node[above right,inner sep=0pt] at (0,0)  {\usebox{0}};
  \node[black] at (0.53\wd0,0.0\ht0) {$t$ [ns]};
  \node[black,rotate=90] at (0.02\wd0,0.5\ht0) {$\mu_{eo}$ [m$^2$/V.s] };
  \node[black] at (0.00\wd0,0.9\ht0) {$\times 10^{-8}$};
  \node[black] at (0.01\wd0,1.0\ht0) {(a) };
  \node[black] at (0.6\wd0,0.68\ht0) {PNP};
  \node[black] at (0.69\wd0,0.62\ht0) {fitted (PNP)};
  \node[black] at (0.62\wd0,0.54\ht0) {LBFT};
  \node[black] at (0.71\wd0,0.47\ht0) {fitted (LBFT)};
\end{tikzpicture}
\begin{tikzpicture}
  \sbox0{\includegraphics[width=2.0in]{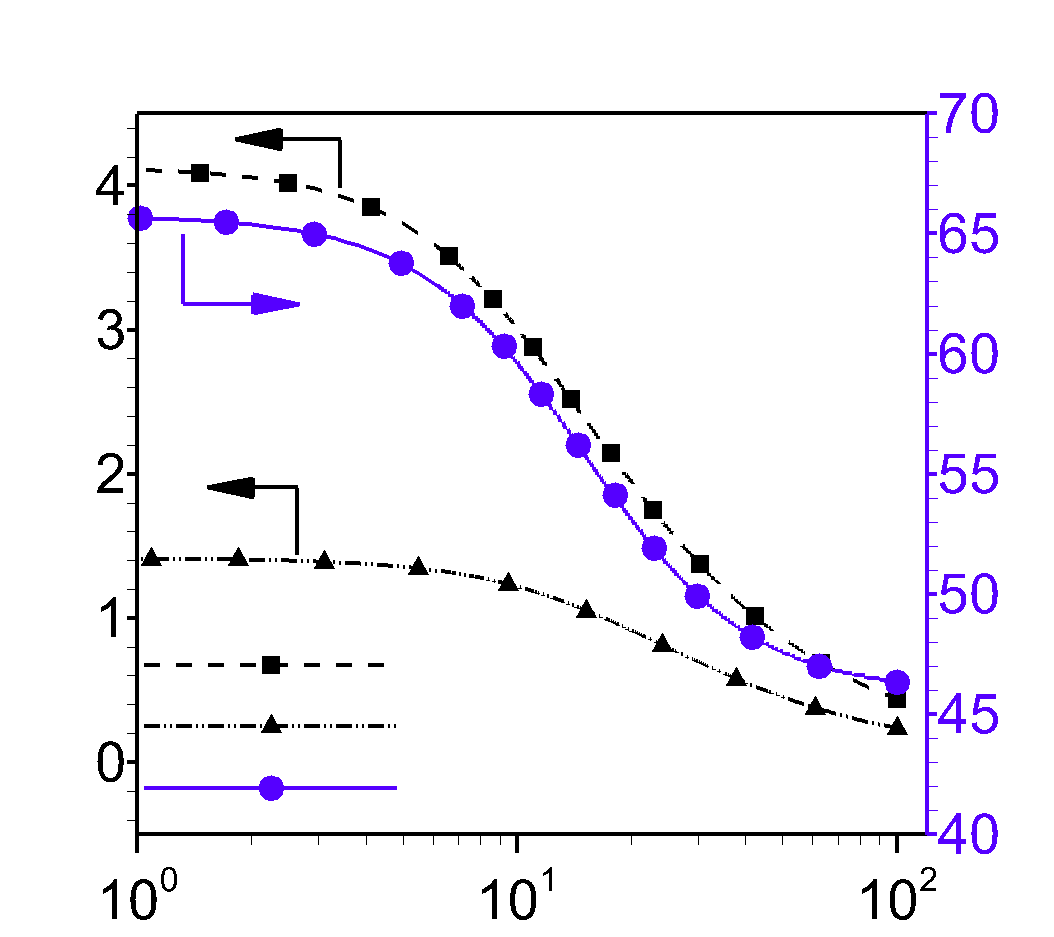} } 
  \node[above right,inner sep=0pt] at (0,0)  {\usebox{0}};
  \node[black] at (0.53\wd0,-0.03\ht0) {$f$ [MHz]};
  \node[black,rotate=90] at (0.02\wd0,0.5\ht0) {$\mu_{eo,rms}$ [m$^2$/V.s] };
  \node[black] at (0.03\wd0,0.9\ht0) {$\times 10^{-8}$};
  \node[blue] at (1.05\wd0,0.55\ht0) {$\frac{(\Delta \mu_{eo})}{\mu_{eo,PNP}}$};
  \node[blue] at (1.0\wd0,0.9\ht0) {[\%]};
  \node[black] at (0.45\wd0,0.3\ht0) {PNP};
  \node[black] at (0.47\wd0,0.23\ht0) {LBFT};
  \node[blue] at (0.62\wd0,0.16\ht0) {dipolar correction};
  \node[black] at (-0.05\wd0,1.0\ht0) {(b) };
\end{tikzpicture}

\caption{(a) Transient electroosmotic mobility following a step voltage, showing the model predictions and the corresponding exponential fits. (b) RMS electroosmotic mobility and the associated dipolar solvent corrections as functions of AC excitation frequency obtained from the fitted transient response. (Color online.)}
\label{fig:7_ACEK}
\end{figure}

The transient evolution of the electroosmotic mobility, in dimensional values, following a step change in surface potential is shown in Fig.~\ref{fig:7_ACEK}(a) for case I of Table~\ref{tab:1Cases}. Similar to the total EDL charge shown in Fig.~\ref{fig:2}(a), the electroosmotic mobility is accurately represented by a single Debye-like exponential relaxation,
\begin{equation}
\mu_{eo}(t)
=
\mu_{eo}^{(\infty)}
\left(
1-e^{-t/\tau_{eo}}
\right),
\end{equation}
where $\tau_{eo}$ denotes the characteristic relaxation time of the coupled EDL--electrohydrodynamic system. For the representative case shown in Fig.~\ref{fig:7_ACEK}(a), the fitted parameters are $\tau_{eo}=9.43$ ns and $\mu_{eo}^{(\infty)}=2.0\!\times\! 10^{-8}$ m$^2$/V\,s for the classical PNP model, and $\tau_{eo}\!=\!14.9$ ns and $\mu_{eo}^{(\infty)}\!=\!5.84 \! \times\! 10^{-8}$ m$^2$/V\,s for the present LBFT model. The longer relaxation time predicted by the LBFT model reflects the slower electrohydrodynamic response arising from the combined effects of dielectric saturation and viscoelectricity.

The characteristic relaxation time obtained from the transient step response can be used to estimate the electroosmotic response under harmonic excitation. Assuming that the mobility is well approximated by a first-order linear relaxation, Duhamel's superposition principle (equivalently, the convolution theorem), widely employed in dielectric relaxation and electrochemical impedance analysis \citep{lazanas2023electrochemical,barsoukov2005impedance,hahn2012heat}, yields the electroosmotic mobility under a sinusoidal surface potential as
\begin{equation}
\mu_{eo}(t)
=
\mu_{eo}^{(\infty)}
\frac{1}
{\sqrt{1+4\pi^2f_{ac}^{\,2}\tau_{eo}^2}}
\sin\!\left(2\pi f_{ac}t-\phi\right),
\qquad
\phi=\tan^{-1}(2\pi f_{ac}\tau_{eo}),
\label{eq:muac}
\end{equation}
where $f_{ac}$ denotes the applied AC frequency. Since the time-average of Eq.~(\ref{eq:muac}) over one cycle is zero, the root-mean-square (RMS) mobility provides a convenient measure of the oscillatory response,
\begin{equation}
\mu_{eo,\mathrm{RMS}}
=
\sqrt{
f_{ac}
\int_{0}^{1/f_{ac}}\!\!\!\!
\mu_{eo}^2(\tilde t)\,
d\tilde t
}
=
\frac{\mu_{eo}^{(\infty)}}
{\sqrt{2\left(1+4\pi^2f_{ac}^{\,2}\tau_{eo}^2\right)}}.
\label{eq:mu_eo_rms}
\end{equation}

The predicted RMS mobility is sensitive to the applied frequency over the characteristic range associated with EDL charging, approximately $1$--$100$ MHz for the present parameters, beyond which the response approaches its asymptotic high-frequency limit [Fig.~\ref{fig:7_ACEK}(b)]. The relative dipolar solvent correction,
\begin{equation}
\Delta\mu_{eo}
=
\frac{\mu_{eo,\mathrm{RMS}}^{\mathrm{(LBFT)}}
-
\mu_{eo,\mathrm{RMS}}^{\mathrm{(PNP)}}}
{\mu_{eo,\mathrm{RMS}}^{\mathrm{(PNP)}}},
\end{equation}
decreases from approximately $65\%$ at low frequencies to an asymptotic value of about $45\%$ above $100$ MHz. This behaviour indicates that transient dipolar solvent effects are most pronounced over the characteristic electrokinetic relaxation timescale, while remaining significant even in the high-frequency limit. As experimental techniques continue to extend into the tens- to hundreds-of-megahertz regime, these results provide a continuum framework for interpreting transient electroosmotic transport under high-frequency forcing.

\section{\label{sec:discussion}Discussion}

Although the present analysis employs the canonical planar electrolytic-cell configuration, its objective is not to reproduce a particular nanopore geometry but to isolate the fundamental coupling between transient EDL charging, dipolar solvent polarization, and electroosmotic forcing. At the microscale, the EDL charging timescale is typically too short for transient electroosmotic flow to be of practical significance. Under nanoscale confinement, however, these timescales become experimentally relevant, especially in modern and emerging nanofluidic systems, making the influence of molecular solvent physics on transient electroosmotic transport an important consideration. Since these mechanisms are governed primarily by the local electric field and interfacial charging dynamics, they are largely independent of the global device geometry during the charging process. Accordingly, the present formulation should be viewed as a generic transient continuum framework that establishes the constitutive influence of dipolar solvent effects, rather than as a geometry-specific nanopore model.

The adopted configuration is motivated by the classical analytical and experimental studies of transient EDL charging, including those used to interpret nanopore charging dynamics \citep{bazant2004diffuse,tivony2018charging}. The transient scaling laws and characteristic frequency transition identified here therefore provide a physically transparent benchmark for extending continuum electrokinetic models to nanoscale fluidics, where transverse EDL evolution under nanoscale confinement plays a critical role in determining the transient fluidic response of the overall device, such as nanopores, nanochannels, and nanoslits. In particular, the predicted frequency-dependent transition in electroosmotic mobility identifies the operating regime over which classical PNP-based descriptions begin to deviate from solvent-informed continuum models. This transition provides both a generic framework for future nanoscale electrokinetic analyses and a basis for experimental validation through extensions of the transient charging measurements of \citet{tivony2018charging}.

Dipolar solvent corrections applied solely to the relative permittivity produce a noticeable deviation of the LBFT model from the MPNP formulation primarily for Case~I. However, when both permittivity and viscosity corrections are incorporated within the LBFT--Stokes framework, departures from the corresponding MPNP--Stokes model arise for Cases~I--III. In contrast, when the dimensionless parameter $\alpha \ll 1$ and either the bulk concentration $\hat{c}_0$ or the applied potential $\hat{V}_0$ is small, as in Cases~IV and~V, the predictions of the LBFT--Stokes model become nearly indistinguishable from those of the traditional MPNP--Stokes formulation. Indicating the criteria for a strong dipolar solvent effect. All-in-all, compared with the constant-property PNP--Stokes model, the LBFT--Stokes formulation systematically predicts reduced electroosmotic mobility. 

The steady-state mobility in Case I is lower by up to 10--20\% solely due to permittivity saturation, whereas the combined dipolar-solvent and viscoelectric effects suppress the mobility by as much as 50--65\% under strong electric-field conditions.
The viscoelectric relation employed here is an empirical constitutive model supported by both direct measurements and molecular dynamics simulations at electric-field strengths comparable to those within nanoconfined electric double layers considered in this work \citep{jin2022direct,zong2016viscosity}. It is therefore interpreted as an effective field-dependent viscosity for continuum electrokinetic modeling rather than as a low-field perturbation expansion.

In the present formulation, the externally imposed quantity is the surface potential, while the surface charge evolves self-consistently during EDL charging and is not prescribed or modulated. The transient electroosmotic response therefore arises from the coupled evolution of ion distributions, field-dependent solvent permittivity, and viscoelectricity, rather than from a time-dependent surface charge alone. In particular, while the transient surface charge and EOF mobility exhibit correlated trends, the substantial reduction in mobility is governed primarily by the field-dependent viscoelectric effect, with dielectric saturation providing an additional contribution.

Direct comparisons with previously reported transient electrokinetic simulations or experiments remain limited, as most existing nanoscale studies have focused on steady-state formulations of dipolar continuum models. Transient electroosmotic behavior is inferred from indirect experimental observables, including ionic current \citep{peng2016electroosmotic,brown2020deconvolution}, streaming current \citep{rheinlaender2025quantifying}, and tracer transport \citep{haywood2014electroosmotic}, interpreted using continuum or hybrid models. In this context, the present continuum framework with non-continuum dipolar solvent effects provides a physically consistent basis for interpreting and predicting transient electrohydrodynamic transport in nanoconfined systems. The predicted frequency dependence establishes a direct link between transient electrokinetic relaxation and experimentally measurable high-frequency electroosmotic transport, with the characteristic transition occurring near the inverse EDL charging time.
The present results establish that transient dipolar solvent corrections are intrinsically linked to EDL charging dynamics through a common characteristic relaxation frequency. This connection provides a unified continuum description of ion transport and solvent polarization in unsteady nanoconfined flows.
As AC electrokinetic experiments continue to extend into the tens- to hundreds-of-megahertz regime, frequency-resolved electroosmotic mobility measurements offer a direct route to validating continuum models incorporating dielectric saturation and viscoelectric effects.

Extension of the present framework to higher-dimensional geometries may reveal additional consequences of dipolar solvent response. In particular, spatial variations of permittivity and charge density can generate tangential Maxwell stresses that drive secondary electrohydrodynamic phenomena, including electroviscous effects arising from coupled ion and fluid transport. Such mechanisms may become especially relevant in nanoconfined flows where strong electric fields coexist with geometric confinement. In nanoscale systems, such as nanopores, nanochannels and nanoscale confinements between electrodes or charged surface, the advective contribution to ionic transport are typically weaker; however, including this effect would enable a fully coupled electrohydrodynamic description and may be pursued in future work. Also, since the Stern layer can play a direct role in transient EDL charging, incorporating a more detailed description of Stern-layer physics within the present continuum framework remains an important direction for future research.

Transient EDL dynamics may also play a role in emerging nanofluidic technologies such as nanopore biosensing and DNA sequencing. During molecular translocation through nanopores, the local ionic environment evolves dynamically in response to both rapid molecular motion and time-dependent electric forcing. Because the characteristic time scale for EDL relaxation can be comparable to molecular transit times, nonequilibrium charge structures may persist and influence ionic current signals. A physically grounded transient EDL framework therefore has the potential to complement data-driven approaches currently used in nanopore signal analysis by providing mechanistic insight into the electrostatic environment surrounding translocating biomolecules \cite{zhao2025nanopore,laucirica2024advances}.

Finally, the LBFT formulation provides a computationally efficient approximation for incorporating dipolar solvent physics into continuum electrokinetic models. In the present implementation, nonuniform but time-invariant permittivity and viscosity fields are employed, consistent with current continuum modeling practices. A more complete formulation with fully time-dependent dipolar coupling, referred to as the dipolar PNP--Stokes (dPNPS) model, is developed separately in \citet{srinivasula2026electrohydrodynamic}. Although the dPNPS model provides a more rigorous description of solvent polarization dynamics, it is computationally more demanding. For a representative simulation on a workstation with a 4~GHz processor and 16~GB RAM, the LBFT--Stokes model requires approximately 15 minutes of computation time, whereas the dPNPS model requires roughly 100 minutes. The LBFT approach therefore offers a practical compromise between physical fidelity and computational efficiency, making it attractive for engineering analyses of transient nanofluidic systems.

\section{\label{sec:conclusions}Conclusions}

This work extends Langevin--Bikerman-type solvent modeling to transient nanoscale electrohydrodynamics. The classical one-dimensional electrolytic cell problem, widely used to study time-dependent PNP formulations for EDL charging with Stern layers and steric effects \cite{bazant2004diffuse,kilic2007bsteric}, is revisited here with the co-evolution of ionic distributions, field-dependent permittivity, and hydrodynamic response under the Stokes flow and weak ionic advection limits. 
The results show that dielectric saturation alone can reduce electroosmotic mobility by up to 10\% relative to standard PNP predictions, while its combined action with the viscoelectric effect leads to reductions as large as 65\%. Although the dominant correction arises from viscoelectricity, the transient evolution of the response depends on the coupled action of both mechanisms.
Within the familiar PNP--Stokes framework, the proposed LB-fitted transient (LBFT) model provides a tractable continuum approach for incorporating dipolar solvent physics into nanoscale electrohydrodynamic simulations. This framework predicts that dipolar solvent physics leaves a distinct frequency-dependent fingerprint on electroosmotic transport, providing a measurable signature for high-frequency AC electrokinetic experiments and predictive modeling. This establishes transient solvent polarization as a continuum-scale mechanism governing unsteady nanofluidic transport relevant to the emerging ultrafast nanofluidic systems.\\

$\quad \quad \quad \quad \quad \quad \quad \quad \quad \quad \quad \quad \quad \quad \quad \quad \quad \quad \quad \quad \quad \quad $ \textbf{DECLARATIONS}\\
No data were created or used in this study. This research received no specific grant from any funding agency. The author declares no competing interests.
\appendix

\section{Appendixes}
\subsection{Steadystate Modeling With Dipolar Solvent}
\begin{figure}
\centering
\begin{tikzpicture}
  \sbox0{\includegraphics[width=2.0in]{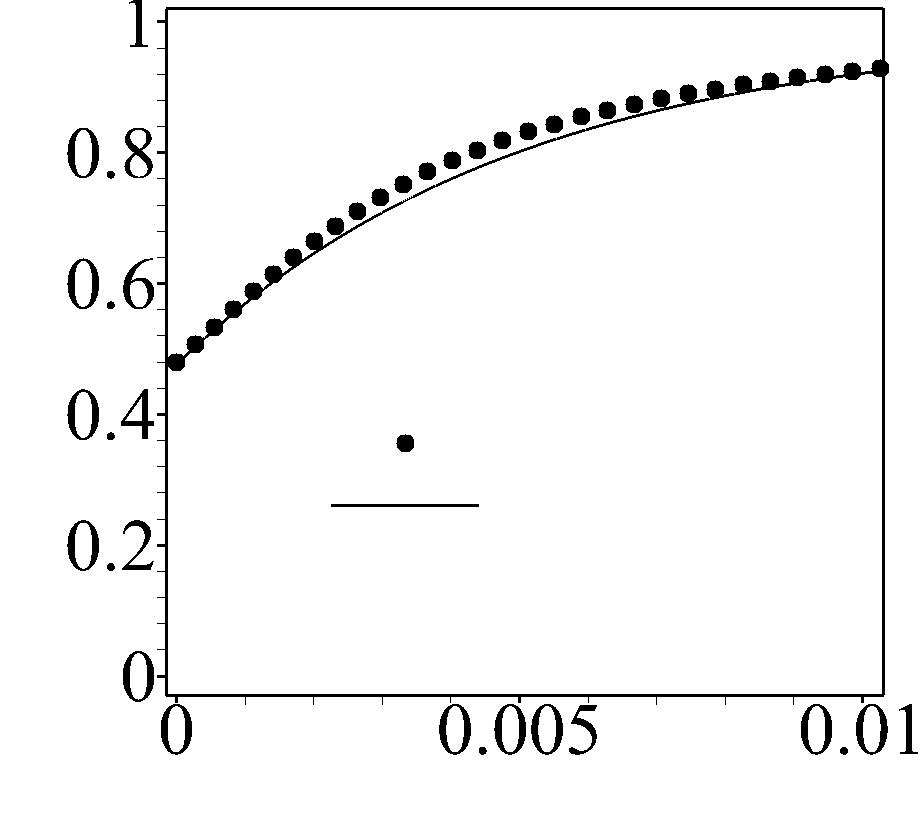} } 
  \node[above right,inner sep=0pt] at (0,0)  {\usebox{0}};
  \node[black] at (0.58\wd0,0.02\ht0) {$\hat{x}-\hat \lambda_s$};
  \node[black,rotate=90] at (0.01\wd0,0.55\ht0) { $\varepsilon_r$/$\varepsilon_{r0}$ };
  \node[black] at (0.01\wd0,1.0\ht0) {(a) };
\node[black] at (0.64\wd0,0.64\ht0) {$A_1 \! \!\left(\!1\!-\!e^{-A_2 \left(\!\frac{x-\lambda_s}{L}+1\!\right)} \!\right)\! +\! A_3$};
\node[black] at (0.71\wd0,0.47\ht0) { LB model};
\node[black] at (0.65\wd0,0.4\ht0) { LBFT};
\end{tikzpicture}
\caption{Spatial profile of permittivity from LBFT and its mathematical fitted function for case I.}
\label{fig:append_LB}
\end{figure}

Langevin-Bikerman (LB) model \cite{gongadze2013spatial} is a steady state model that incorporates dipolar solvent effects via nonuniform permittivity given by,
\begin{align}\label{eq:perm_LB}
    \varepsilon_r &=1+ n_{0w}n_s\frac{p_0}{\varepsilon_0}\frac{\mathcal{F}}{E\mathcal{H}}\\
    \mathcal{F}&= \left(coth(p_0E\beta)-\frac{1}{p_0E\beta} \right) \frac{sinh(p_0E\beta)}{p_0E\beta}\\
    \mathcal{H} &= 2n_0 cosh(e_0\psi \beta) + n_{0w}\frac{sinh(p_0E\beta)}{p_0E\beta},
\end{align}
into PB model as
\begin{align}
    \nabla\cdot(\varepsilon_0\varepsilon_r \nabla \psi) &= -\rho_i^{LB}\\
    \rho_i^{LB} &= -2e_0n_0n_s \frac{sinh(e_0\beta\psi)}{\mathcal{H}}.
\end{align}
Here $n_0$ denotes the bulk ionic concentration, $n_{0w}$ the bulk solvent concentration, $p_0$ the effective solvent dipole moment, and $\beta=(k_BT)^{-1}$. Following classical electrochemical cell models, we consider one half of a symmetric electrolyte domain bounded by a Stern layer adjacent to the electrode. The LB equations are solved in the diffuse layer, $-(1-\epsilon\delta_s)\le\hat{x}\le0$, while a Laplace equation governs the Stern layer, $-1\le\hat{x}<-(1-\epsilon\delta_s)$. A fixed applied potential $\hat{\psi}\!=\!-v$ is prescribed at the electrode surface, and for $\epsilon\ll1$ the bulk boundary is taken to be electrically neutral, $\nabla\hat{\psi}=0$, ensuring an unperturbed far-field electrolyte.

\begin{table}[h]
    \centering
    \caption{Fitting parameters of the exponential-form approximation for the spatially varying relative permittivity in Eq.~\ref{eq:dimen_Poisson}, corresponding to the parametric cases listed in Table~\ref{tab:1Cases}.}
    \label{tab:permitt_coeff}
    \begin{tabular}{lrrr}
        \hline
        Generic form & $A_1$ & $A_2$ & $A_3$ \\
        \hline
        case-1 & $0.521$& $200$& $0.480$\\
        case-2 & $0.044$& $33.33$& $0.956$\\
        case-3 & $0.044$& $33.33$& $0.956$\\
        case-4 & $7\times10^{-5}$& $2.99$& $0.999$\\
        case-5 & $2.7\times10^{-4}$& $20.0$& $0.999$\\
        \hline
    \end{tabular}
\end{table}

Spatially nonuniform permittivity of case I obtained from the LB model is shown in Figure~\ref{fig:append_LB}. Fitted exponential profiles (of the form indicated in Eq. \ref{eq:dimen_Poisson}) for the different cases are listed in Table \ref{tab:permitt_coeff}. As counterions move to the surface and increase the electric field and its gradient, solvent concentration decreases, and the polarization density decreases\cite{gongadze2013spatial}. This results in lower permittivity, as seen in the figure, estimated by the LB model \citep[see][eq. 36--38]{gongadze2013spatial}. An exponential profile fitted to the LB solution is also shown here, which is used in the LBFT model for this parametric case.
For $\alpha\ll 1$, $\mathcal{F}\approx \alpha/3$ and $\mathcal{H}\approx n_{0w}(1+(2n_0/n_{0w})cosh(\tilde{\psi}))$, for $\tilde{\psi}=\psi/(k_BT/e_0)$. Hence the above permittivity expression converges to a uniform value of the MPNP limit.

\bibliography{apssamp}

@article{iglivc2010excluded,
  title={Excluded volume effect and orientational ordering near charged surface in solution of ions and Langevin dipoles},
  author={Igli{\v{c}}, Ale{\v{s}} and Gongadze, Ekaterina and Bohinc, Klemen},
  journal={Bioelectrochemistry},
  volume={79},
  number={2},
  pages={223--227},
  year={2010},
  publisher={Elsevier}
}

@article{pandey2021impact,
  title={Impact of finite ion size, Born energy difference and dielectric decrement on the electroosmosis of multivalent ionic mixtures in a nanotube},
  author={Pandey, Doyel and Bhattacharyya, Somnath},
  journal={Colloids and Surfaces A: Physicochemical and Engineering Aspects},
  volume={610},
  pages={125905},
  year={2021},
  publisher={Elsevier}
}

@article{pandey2022effects,
  title={Effects of membrane polarization, steric repulsion and ion-solvent interactions on electroosmosis through a conical nanopore},
  author={Pandey, Doyel and Bhattacharyya, Somnath},
  journal={Applied Mathematical Modelling},
  volume={111},
  pages={471--485},
  year={2022},
  publisher={Elsevier}
}

@article{gongadze2013spatial,
  title={Spatial variation of permittivity of an electrolyte solution in contact with a charged metal surface: a mini review},
  author={Gongadze, Ekaterina and van Rienen, Ursula and Kralj-Igli{\v{c}}, Veronika and Igli{\v{c}}, Ale{\v{s}}},
  journal={Computer methods in biomechanics and biomedical engineering},
  volume={16},
  number={5},
  pages={463--480},
  year={2013},
  publisher={Taylor \& Francis}
}

@article{abrashkin2007dipolar,
  title={Dipolar Poisson-Boltzmann equation: ions and dipoles close to charge interfaces},
  author={Abrashkin, Ariel and Andelman, David and Orland, Henri},
  journal={Physical review letters},
  volume={99},
  number={7},
  pages={077801},
  year={2007},
  publisher={APS}
}

@article{kilic2007bsteric,
  title={Steric effects in the dynamics of electrolytes at large applied voltages. II. Modified Poisson-Nernst-Planck equations},
  author={Kilic, Mustafa Sabri and Bazant, Martin Z and Ajdari, Armand},
  journal={Physical Review E—Statistical, Nonlinear, and Soft Matter Physics},
  volume={75},
  number={2},
  pages={021503},
  year={2007},
  publisher={APS}
}

@article{borukhov1997steric,
  title={Steric effects in electrolytes: A modified Poisson-Boltzmann equation},
  author={Borukhov, Itamar and Andelman, David and Orland, Henri},
  journal={Physical review letters},
  volume={79},
  number={3},
  pages={435},
  year={1997},
  publisher={APS}
}

@article{bazant2004diffuse,
  title={Diffuse-charge dynamics in electrochemical systems},
  author={Bazant, Martin Z and Thornton, Katsuyo and Ajdari, Armand},
  journal={Physical Review E—Statistical, Nonlinear, and Soft Matter Physics},
  volume={70},
  number={2},
  pages={021506},
  year={2004},
  publisher={APS}
}

@article{pennathur2005electrokinetic,
  title={Electrokinetic transport in nanochannels. 2. Experiments},
  author={Pennathur, Sumita and Santiago, Juan G},
  journal={Analytical chemistry},
  volume={77},
  number={21},
  pages={6782--6789},
  year={2005},
  publisher={ACS Publications}
}

@article{mehta2025arresting,
  title={Arresting of Viscoelectric Effect Modulated Flow Reduction in Nanochannels with Imposed Temperature Gradients},
  author={Mehta, Sumit Kumar and Biswas, Gautam and Mondal, Pranab Kumar},
  journal={Langmuir},
  volume={41},
  number={30},
  pages={19754--19767},
  year={2025},
  publisher={ACS Publications}
}

@article{hsu2016electrokinetics,
  title={Electrokinetics of the silica and aqueous electrolyte solution interface: Viscoelectric effects},
  author={Hsu, Wei-Lun and Daiguji, Hirofumi and Dunstan, David E and Davidson, Malcolm R and Harvie, Dalton JE},
  journal={Advances in Colloid and Interface Science},
  volume={234},
  pages={108--131},
  year={2016},
  publisher={Elsevier}
}

@article{saurabh2023mathematical,
  title={Mathematical and computational modeling of electrohydrodynamics through a nanochannel},
  author={Saurabh, Kumar and Solovchuk, Maxim},
  journal={AIP Advances},
  volume={13},
  number={1},
  year={2023},
  publisher={AIP Publishing}
}

@article{jin2022direct,
  title={Direct measurement of the viscoelectric effect in water},
  author={Jin, Di and Hwang, Yongyun and Chai, Liraz and Kampf, Nir and Klein, Jacob},
  journal={Proceedings of the National Academy of Sciences},
  volume={119},
  number={1},
  pages={e2113690119},
  year={2022},
  publisher={National Academy of Sciences}
}

@article{blankenburg2025nanoscale,
  title={Nanoscale AC Electroosmotic Flow and the Frequency--Size Scaling Observed beyond the Charge Relaxation Regime},
  author={Blankenburg, Gerhard and Hern{\'a}ndez-Alp{\'\i}zar, Huberth and Lesser-Rojas, Leonardo and Chou, Chia-Fu},
  journal={Nano Letters},
  year={2025},
  publisher={ACS Publications}
}

@article{choi2024electroconvective,
  title={Electroconvective instability at the surface of one-dimensionally patterned ion exchange membranes},
  author={Choi, Jinwoong and Cho, Myeonghyeon and Shin, Joonghan and Kwak, Rhokyun and Kim, Bumjoo},
  journal={Journal of Membrane Science},
  volume={691},
  pages={122256},
  year={2024},
  publisher={Elsevier}
}

@article{de2018confined,
  title={Confined electroconvective vortices at structured ion exchange membranes},
  author={De Valenca, Joeri and Jogi, Morten and Wagterveld, R Martijn and Karatay, Elif and Wood, Jeffery A and Lammertink, Rob GH},
  journal={Langmuir},
  volume={34},
  number={7},
  pages={2455--2463},
  year={2018},
  publisher={ACS Publications}
}

@article{tian2021theory,
  title={Theory of shock electrodialysis I: Water dissociation and electrosmotic vortices},
  author={Tian, Huanhuan and Alkhadra, Mohammad A and Bazant, Martin Z},
  journal={Journal of Colloid and Interface Science},
  volume={589},
  pages={605--615},
  year={2021},
  publisher={Elsevier}
}

@article{sin2018influence,
  title={Influence of solvent polarization and non-uniform ionic size on electrokinetic transport in a nanochannel},
  author={Sin, Jun-Sik and Kim, Nam-Hyok and Kim, Chol-Ho and Jang, Yong-Man},
  journal={Microfluidics and Nanofluidics},
  volume={22},
  pages={1--13},
  year={2018},
  publisher={Springer}
}

@article{sinha2016effect,
  title={Effect of solvent polarization on electroosmotic transport in a nanofluidic channel},
  author={Sinha, Shayandev and Myers, Lucas and Das, Siddhartha},
  journal={Microfluidics and Nanofluidics},
  volume={20},
  pages={1--14},
  year={2016},
  publisher={Springer}
}

@article{srinivasula2025dipolar,
  title={Dipolar solvent corrections in nanopore electroosmotic flow with different surface electrostatic conditions},
  author={Srinivasula, Pramodt},
  journal={Journal of Electrostatics},
  volume={138},
  pages={104201},
  year={2025},
  publisher={Elsevier}
}

@article{liu2018modified,
  title={Modified Poisson--Nernst--Planck model with accurate Coulomb correlation in variable media},
  author={Liu, Pei and Ji, Xia and Xu, Zhenli},
  journal={SIAM Journal on Applied Mathematics},
  volume={78},
  number={1},
  pages={226--245},
  year={2018},
  publisher={SIAM}
}

@article{wei2024nanopore,
  title={Nanopore-based sensors for DNA sequencing: a review},
  author={Wei, Jiangtao and Hong, Hao and Wang, Xing and Lei, Xin and Ye, Minjie and Liu, Zewen},
  journal={Nanoscale},
  volume={16},
  number={40},
  pages={18732--18766},
  year={2024},
  publisher={Royal Society of Chemistry}
}

@article{he2011controlling,
  title={Controlling DNA translocation through gate modulation of nanopore wall surface charges},
  author={He, Yuhui and Tsutsui, Makusu and Fan, Chun and Taniguchi, Masateru and Kawai, Tomoji},
  journal={ACS nano},
  volume={5},
  number={7},
  pages={5509--5518},
  year={2011},
  publisher={ACS Publications}
}

@article{gui2024ion,
  title={Ion Transport in Dipolar Medium I: A Local Dielectric Poisson--Nernst--Planck/Poisson--Boltzmann Model},
  author={Gui, Sheng and Lu, Benzhuo and Yu, Weilin},
  journal={SIAM Journal on Applied Mathematics},
  volume={84},
  number={5},
  pages={2110--2131},
  year={2024},
  publisher={SIAM}
}

@article{liu2020molecular,
  title={Molecular mean-field theory of ionic solutions: A Poisson-Nernst-Planck-Bikerman model},
  author={Liu, Jinn-Liang and Eisenberg, Bob},
  journal={Entropy},
  volume={22},
  number={5},
  pages={550},
  year={2020},
  publisher={MDPI}
}

@article{di2025validity,
  title={Validity of approximated expressions for electro-osmotic flow in nanopores evaluated by continuum electrohydrodynamics and atomistic simulations},
  author={Di Muccio, Giovanni and Gargano, Simone and Iacoviello, Domingo Francesco and della Rocca, Blasco Morozzo and Chinappi, Mauro},
  journal={Flow},
  volume={5},
  pages={E28},
  year={2025},
  publisher={Cambridge University Press}
}

@article{andrade1951effect,
  title={The effect of an electric field on the viscosity of liquids. II},
  author={Andrade, Edward Neville Da Costa and Dodd, C},
  journal={Proceedings of the Royal Society of London. Series A. Mathematical and Physical Sciences},
  volume={204},
  number={1079},
  pages={449--464},
  year={1951},
  publisher={The Royal Society London}
}

@article{andrade1946effect,
  title={The effect of an electric field on the viscosity of liquids},
  author={Andrade, Edward Neville Da Costa and Dodd, C},
  journal={Proceedings of the Royal Society of London. Series A. Mathematical and Physical Sciences},
  volume={187},
  number={1010},
  pages={296--337},
  year={1946},
  publisher={The Royal Society London}
}

@article{hunter1978viscoelectric,
  title={Viscoelectric coefficient for water},
  author={Hunter, Robert J and Leyendekkers, JV},
  journal={Journal of the Chemical Society, Faraday Transactions 1: Physical Chemistry in Condensed Phases},
  volume={74},
  pages={450--455},
  year={1978},
  publisher={Royal Society of Chemistry}
}

@article{lyklema1961interpretation,
  title={On the interpretation of electrokinetic potentials},
  author={Lyklema, J and Overbeek, J Th G},
  journal={Journal of Colloid Science},
  volume={16},
  number={5},
  pages={501--512},
  year={1961},
  publisher={Elsevier}
}

@article{fumagalli2018anomalously,
  title={Anomalously low dielectric constant of confined water},
  author={Fumagalli, Laura and Esfandiar, Ali and Fabregas, Rene and Hu, Sheng and Ares, Pablo and Janardanan, Amritha and Yang, Qian and Radha, Boya and Taniguchi, Takashi and Watanabe, Kenji and others},
  journal={Science},
  volume={360},
  number={6395},
  pages={1339--1342},
  year={2018},
  publisher={American Association for the Advancement of Science}
}

@article{monet2021nonlocal,
  title={Nonlocal dielectric response of water in nanoconfinement},
  author={Monet, Geoffrey and Bresme, Fernando and Kornyshev, Alexei and Berthoumieux, H{\'e}l{\`e}ne},
  journal={Physical Review Letters},
  volume={126},
  number={21},
  pages={216001},
  year={2021},
  publisher={APS}
}

@article{zheng2006surfaces,
  title={Surfaces and interfacial water: evidence that hydrophilic surfaces have long-range impact},
  author={Zheng, Jian-ming and Chin, Wei-Chun and Khijniak, Eugene and Khijniak Jr, Eugene and Pollack, Gerald H},
  journal={Advances in colloid and interface science},
  volume={127},
  number={1},
  pages={19--27},
  year={2006},
  publisher={Elsevier}
}

@article{ratschow2022resonant,
  title={Resonant nanopumps: ac gate voltages in conical nanopores induce directed electrolyte flow},
  author={Ratschow, Aaron D and Pandey, Doyel and Liebchen, Benno and Bhattacharyya, Somnath and Hardt, Steffen},
  journal={Physical Review Letters},
  volume={129},
  number={26},
  pages={264501},
  year={2022},
  publisher={APS}
}

@article{he2011gate,
  title={Gate manipulation of DNA capture into nanopores},
  author={He, Yuhui and Tsutsui, Makusu and Fan, Chun and Taniguchi, Masateru and Kawai, Tomoji},
  journal={ACS nano},
  volume={5},
  number={10},
  pages={8391--8397},
  year={2011},
  publisher={ACS Publications}
}

@article{saville1997electrohydrodynamics,
  title={Electrohydrodynamics: the Taylor-Melcher leaky dielectric model},
  author={Saville, DA1435033},
  journal={Annual review of fluid mechanics},
  volume={29},
  number={1},
  pages={27--64},
  year={1997},
  publisher={Annual Reviews 4139 El Camino Way, PO Box 10139, Palo Alto, CA 94303-0139, USA}
}

@article{ladiges2021discrete,
  title={Discrete ion stochastic continuum overdamped solvent algorithm for modeling electrolytes},
  author={Ladiges, Daniel R and Nonaka, A and Klymko, K and Moore, GC and Bell, JB and Carney, SP and Garcia, AL and Natesh, SR and Donev, A},
  journal={Physical Review Fluids},
  volume={6},
  number={4},
  pages={044309},
  year={2021},
  publisher={APS}
}

@article{sahu2024ion,
  title={Ion steric interactions and electrostatic correlations on electro-osmotic flow in charged nanopores with multivalent electrolytes},
  author={Sahu, Shubhra and Mondal, Bapan and Bhattacharyya, Somnath},
  journal={Physical Review Fluids},
  volume={9},
  number={7},
  pages={074201},
  year={2024},
  publisher={APS}
}

@article{wang2025field,
  title={Field-effect nanofluidic memristor},
  author={Wang, Wei and Ma, Yu and Liang, Yizheng},
  journal={Physics of Fluids},
  volume={37},
  number={8},
  year={2025},
  publisher={AIP Publishing}
}

@article{dolatshahi2025salinity,
  title={Salinity gradient energy harvesting via tuned ionic nanotransistor geometries},
  author={Dolatshahi, Reza and Khatibi, Mahdi and Ashrafizadeh, Seyed Nezameddin},
  journal={Physics of Fluids},
  volume={37},
  number={10},
  year={2025},
  publisher={AIP Publishing}
}

@article{peng2025ionic,
  title={Ionic current rectification properties of power-law fluids in a nanochannel with non-uniform surface charge density},
  author={Peng, Li and Hao, Yu and Liu, Runxin and Zhao, Zhengyang and Li, Jie},
  journal={Physics of Fluids},
  volume={37},
  number={4},
  year={2025},
  publisher={AIP Publishing}
}

@article{mehta2023viscoelectric,
  title={Viscoelectric effect on the chemiosmotic flow in charged soft nanochannels},
  author={Mehta, Sumit Kumar and Mondal, Pranab Kumar},
  journal={Physics of Fluids},
  volume={35},
  number={11},
  year={2023},
  publisher={AIP Publishing}
}

@article{melnikov2024ionic,
  title={Ionic current blockade in a nanopore due to an ellipsoidal particle},
  author={Melnikov, Dmitriy V and Barker, Nelson R and Gracheva, Maria E},
  journal={Physical Review E},
  volume={110},
  number={3},
  pages={034403},
  year={2024},
  publisher={APS}
}

@article{laucirica2024advances,
  title={Advances in nanofluidic field-effect transistors: external voltage-controlled solid-state nanochannels for stimulus-responsive ion transport and beyond},
  author={Laucirica, Gregorio and Toum-Terrones, Yamili and Cay{\'o}n, VM and Toimil-Molares, ME and Azzaroni, Omar and Marmisoll{\'e}, Waldemar Alejandro},
  journal={Physical Chemistry Chemical Physics},
  volume={26},
  number={14},
  pages={10471--10493},
  year={2024},
  publisher={Royal Society of Chemistry}
}

@article{zhao2025nanopore,
  title={Nanopore toward Genuine Single-Molecule Sensing: Molecular Ping-Pong Technology},
  author={Zhao, Xinjia and Zhang, Yahui and Qing, Guangyan},
  journal={Nano Letters},
  volume={25},
  number={10},
  pages={3692--3706},
  year={2025},
  publisher={ACS Publications}
}

@article{haywood2014electroosmotic,
  title={Electroosmotic flow in nanofluidic channels},
  author={Haywood, Daniel G and Harms, Zachary D and Jacobson, Stephen C},
  journal={Analytical chemistry},
  volume={86},
  number={22},
  pages={11174--11180},
  year={2014},
  publisher={ACS Publications}
}

@article{peng2016electroosmotic,
  title={Electroosmotic flow in single PDMS nanochannels},
  author={Peng, Ran and Li, Dongqing},
  journal={Nanoscale},
  volume={8},
  number={24},
  pages={12237--12246},
  year={2016},
  publisher={Royal Society of Chemistry}
}

@article{rheinlaender2025quantifying,
  title={Quantifying and Utilizing Electroosmotic Flow for Mechanical Measurements with the Scanning Ion Conductance Microscope},
  author={Rheinlaender, Johannes and Schäffer, Tilman E},
  journal={Analytical Chemistry},
  volume={97},
  number={41},
  pages={22541--22547},
  year={2025},
  publisher={ACS Publications}
}

@article{brown2020deconvolution,
  title={Deconvolution of electroosmotic flow in hysteresis ion transport through single asymmetric nanopipettes},
  author={Brown, Warren and Li, Yan and Yang, Ruoyu and Wang, Dengchao and Kvetny, Maksim and Zheng, Hui and Wang, Gangli},
  journal={Chemical science},
  volume={11},
  number={23},
  pages={5950--5958},
  year={2020},
  publisher={Royal Society of Chemistry}
}

@article{alosious2026study,
  title={Study of the molecular level mechanism of nanoscale alternating current electrohydrodynamic flow},
  author={Alosious, Sobin and Antaw, Fiach and Trau, Matt and Tee, Shern R and Searles, Debra J},
  journal={International Communications in Heat and Mass Transfer},
  volume={173},
  pages={110890},
  year={2026},
  publisher={Elsevier}
}

@article{coquinot2026electron,
  title={Electron--electrolyte coupling in AC transport through nanofluidic channels},
  author={Coquinot, Baptiste and Liz{\'e}e, Mathieu and Bocquet, Lyd{\'e}ric and Kavokine, Nikita},
  journal={The Journal of Chemical Physics},
  volume={164},
  number={13},
  year={2026},
  publisher={AIP Publishing}
}

@article{shiddiky2014molecular,
  title={Molecular nanoshearing: An innovative approach to shear off molecules with AC-induced nanoscopic fluid flow},
  author={Shiddiky, Muhammad JA and Vaidyanathan, Ramanathan and Rauf, Sakandar and Tay, Zhikai and Trau, Matt},
  journal={Scientific reports},
  volume={4},
  number={1},
  pages={3716},
  year={2014},
  publisher={Nature Publishing Group UK London}
}

@article{li2025high,
  title={High-frequency supercapacitors surpassing dynamic limit of electrical double layer effects},
  author={Li, Zhangshanhao and Xu, Minghao and Xia, Yier and Yan, Ziyun and Dai, Jianyou and Hu, Bingmeng and Feng, Haizhao and Xu, Sixing and Wang, Xiaohong},
  journal={Nature Communications},
  volume={16},
  number={1},
  pages={3704},
  year={2025},
  publisher={Nature Publishing Group UK London}
}

@article{shekar2016measurement,
  title={Measurement of DNA translocation dynamics in a solid-state nanopore at 100 ns temporal resolution},
  author={Shekar, Siddharth and Niedzwiecki, David J and Chien, Chen-Chi and Ong, Peijie and Fleischer, Daniel A and Lin, Jianxun and Rosenstein, Jacob K and Drndic, Marija and Shepard, Kenneth L},
  journal={Nano letters},
  volume={16},
  number={7},
  pages={4483--4489},
  year={2016},
  publisher={ACS Publications}
}

@article{rosenstein2012integrated,
  title={Integrated nanopore sensing platform with sub-microsecond temporal resolution},
  author={Rosenstein, Jacob K and Wanunu, Meni and Merchant, Christopher A and Drndic, Marija and Shepard, Kenneth L},
  journal={Nature methods},
  volume={9},
  number={5},
  pages={487--492},
  year={2012},
  publisher={Nature Publishing Group US New York}
}

@article{kavokine2021fluids,
  title={Fluids at the nanoscale: from continuum to subcontinuum transport},
  author={Kavokine, Nikita and Netz, Roland R and Bocquet, Lyd{\'e}ric},
  journal={Annual Review of Fluid Mechanics},
  volume={53},
  number={1},
  pages={377--410},
  year={2021},
  publisher={Annual Reviews}
}

@article{abdelghani2025resonant,
  title={Resonant osmotic diodes for voltage-induced water filtration across composite membranes},
  author={Abdelghani-Idrissi, Soufiane and Ries, Lucie and Monet, Geoffrey and Perez-Carvajal, Javier and Pilo, Zacharie and Sarnikowski, Paulina and Siria, Alessandro and Bocquet, Lyd{\'e}ric},
  journal={Nature Materials},
  volume={24},
  number={7},
  pages={1109--1115},
  year={2025},
  publisher={Nature Publishing Group UK London}
}

@article{lazanas2023electrochemical,
  title={Electrochemical impedance spectroscopy─ a tutorial},
  author={Lazanas, Alexandros Ch and Prodromidis, Mamas I},
  journal={ACS measurement science au},
  volume={3},
  number={3},
  pages={162--193},
  year={2023},
  publisher={ACS Publications}
}

@book{barsoukov2005impedance,
  title={Impedance spectroscopy theory, experiment, and applications},
  author={Barsoukov, Evgenij},
  year={2005},
  publisher={Wiley Online Library}
}

@book{hahn2012heat,
  title={Heat conduction},
  author={Hahn, David W and {\"O}zisik, M Necati},
  year={2012},
  publisher={John Wiley \& Sons}
}

@article{maurel2022operando,
  title={Operando AC in-plane impedance spectroscopy of electrodes for energy storage systems},
  author={Maurel, Victor and Brousse, Kevin and Mathis, Tyler S and Perju, Audrey and Taberna, Pierre-Louis and Simon, Patrice},
  journal={Journal of The Electrochemical Society},
  volume={169},
  number={12},
  pages={120510},
  year={2022},
  publisher={IOP Publishing}
}

@article{chapuis2025boosting,
  title={Boosting large scale capacitive harvesting of osmotic power by dynamic matching of ion exchange kinetics},
  author={Chapuis, Nicolas and Bocquet, Lyd{\'e}ric},
  journal={Sustainable Energy \& Fuels},
  volume={9},
  number={8},
  pages={2087--2097},
  year={2025},
  publisher={Royal Society of Chemistry}
}

@article{perkin2006long,
  title={Long-range attraction between charge-mosaic surfaces across water},
  author={Perkin, Susan and Kampf, Nir and Klein, Jacob},
  journal={Physical review letters},
  volume={96},
  number={3},
  pages={038301},
  year={2006},
  publisher={APS}
}

@article{srinivasula2026electrohydrodynamic,
  title={Electrohydrodynamic Stresses from Hydrogen-Bond Network Dynamics in Water},
  author={Srinivasula, Pramodt},
  journal={arXiv preprint arXiv:2603.12941},
  year={2026}
}

@article{palaia2025charging,
  title={Charging dynamics of electric double-layer nanocapacitors in mean field},
  author={Palaia, Ivan and Asta, Adelchi J and Dutta, Megh and Warren, Patrick B and Rotenberg, Benjamin and Trizac, Emmanuel},
  journal={Physical Review Letters},
  volume={135},
  number={14},
  pages={148002},
  year={2025},
  publisher={APS}
}

@article{ratschow2025convection,
  title={Convection can enhance the capacitive charging of porous electrodes},
  author={Ratschow, Aaron D and Wagner, Alexander J and Janssen, Mathijs and Hardt, Steffen},
  journal={Proceedings of the National Academy of Sciences},
  volume={122},
  number={50},
  pages={e2504322122},
  year={2025},
  publisher={National Academy of Sciences}
}

@article{zong2016viscosity,
  title={Viscosity of water under electric field: Anisotropy induced by redistribution of hydrogen bonds},
  author={Zong, Diyuan and Hu, Han and Duan, Yuanyuan and Sun, Ying},
  journal={The Journal of Physical Chemistry B},
  volume={120},
  number={21},
  pages={4818--4827},
  year={2016},
  publisher={ACS Publications}
}

@article{lan2016voltage,
  title={Voltage-rectified current and fluid flow in conical nanopores},
  author={Lan, Wen-Jie and Edwards, Martin A and Luo, Long and Perera, Rukshan T and Wu, Xiaojian and Martin, Charles R and White, Henry S},
  journal={Accounts of chemical research},
  volume={49},
  number={11},
  pages={2605--2613},
  year={2016},
  publisher={ACS Publications}
}

@article{tivony2018charging,
  title={Charging dynamics of an individual nanopore},
  author={Tivony, Ran and Safran, Sam and Pincus, Philip and Silbert, Gilad and Klein, Jacob},
  journal={Nature communications},
  volume={9},
  number={1},
  pages={4203},
  year={2018},
  publisher={Nature Publishing Group UK London}
}

@article{cui2004electroosmotic,
  title={Electroosmotic flow in nanoscale parallel-plate channels: molecular simulation study and comparison with classical Poisson--Boltzmann theory},
  author={Cui, ST and Cochran, HD},
  journal={Molecular Simulation},
  volume={30},
  number={5},
  pages={259--266},
  year={2004},
  publisher={Taylor \& Francis}
}

\end{document}